\documentclass[12pt]{article} 
\usepackage{latexsym} 
\def\a{\alpha} 
\def\b{\beta} 
\def\g{\gamma}  
\def\d{\delta}  
\def\e{\eta} 
\def\h{\eta}

\def\m{\mu}  
\def\n{\nu}  
\def\r{\rho}  
\def\o{\omega} 
\def\s{\sigma}

\def\e{\varepsilon} 
\def\cA{{\cal A}}
 
\def\pa{\partial}
 
\def\ll{\left} 
\def\rr{\right} 
\def\be{\begin{equation}}  
\def\ee{\end{equation}}  
\def\beq{\begin{eqnarray}}  
\def\eeq{\end{eqnarray}}  
\def\nn{\nonumber}

\newcommand{\bqn}{\begin{eqnarray}}\newcommand{\eqn}{\end{eqnarray}}
\newtheorem{theorem}{Theorem}[subsection] 
\newtheorem{lemma}{Lemma}[subsection]

\begin{document} 
\begin{titlepage} 
\begin{centering} 
 
{\huge {\bf Inconsistency of interacting, multi-graviton theories}} 
 
\vspace{3cm} 
 
{\Large Nicolas Boulanger$^{a,}$\footnote{``Chercheur F.R.I.A.", Belgium},  
Thibault Damour$^b$, Leonardo Gualtieri$^a$ and 
Marc Henneaux$^{a,c}$} \\ 
\vspace{.4cm} 
$^a$ Physique Th\'eorique et Math\'ematique,  Universit\'e Libre 
de Bruxelles,  C.P. 231, B-1050, Bruxelles, Belgium      \\ 
\vspace{.2cm} 
$^b$ Institut des Hautes Etudes Scientifiques,  35, route de 
Chartres,  F-91440 Bures-sur-Yvette, France \\ 
\vspace{.2cm} 
$^c$ Centro de Estudios Cient\'{\i}ficos, Casilla 1469, Valdivia, Chile 
\vspace{1cm} \\ 
 
{\tt nboulang@ulb.ac.be, 
damour@ihes.fr, Leonardo.Gualtieri@ulb.ac.be, henneaux@ulb.ac.be}

\vspace{2cm} 
 
\end{centering} 
 
\begin{abstract} 
We investigate, in any spacetime dimension 
$\geq 3$,   the problem of consistent couplings for a 
(finite or infinite) collection 
of massless, spin-2 fields described, in the free limit, by a sum 
of Pauli-Fierz actions. 
We show that there is no consistent (ghost-free) coupling, with at most
two derivatives of the fields, that can mix the various 
``gravitons''. In other words, there are no Yang-Mills-like spin-2
theories.  The only possible deformations are given by a 
sum (or integral) of individual Einstein-Hilbert actions. 
The impossibility of cross-couplings subsists in the 
presence of scalar matter.   
Our approach is based on the BRST-based deformation point of view 
and uses results on the so-called ``characteristic cohomology'' for 
massless spin-2 fields which are explained in detail.  
\end{abstract}

\vfill           
\end{titlepage}  
 
\section{Introduction} 
\setcounter{equation}{0} 
\setcounter{theorem}{0} 
\setcounter{lemma}{0}

A striking feature of the interactions observed in Nature is that most 
of them (if we weigh them by the number of helicity states) are described
by nonlinearly interacting {\it multiplets} of massless spin-1 fields,
i.e., by
Yang-Mills' theory. By contrast, the gravitational interaction 
(Einstein's theory) involves 
only a {\it single} massless spin-two field. In this paper 
we shall show that there is a compelling theoretical reason underlying this
fact: there exists no consistent (in particular, ghost-free) theory 
involving  (finite or infinite) multiplets of interacting massless spin-2 fields.
In other words, there exists no spin-2 analog of Yang-Mills' theory.
This no-go result gives a new argument (besides the usual one based on
the problems of
having particles of spin $ > 2$) for ruling out $ N > 8$ 
extended supergravity theories, since these would involve gravitons
of different types.  

It was shown by Pauli and Fierz \cite{Fierz:1939ix} 
that there is a unique, consistent\footnote{All over this paper, we follow
the standard
field theory tenets which tell us that ``consistent'' theories
should be free of: negative-energy (ghost) propagating
excitations, algebraic
inconsistencies among field equations, discontinuities in the 
degree-of-freedom content, etc.}
action describing a pure spin-2,
massless field. This action happens to be the linearized 
Einstein action. Therefore, 
the action for a collection $\{h_{\mu \nu}^a\}$ 
of $N$ {\it non-interacting}, massless spin-2 fields in spacetime dimension
$n$ ($a = 1, \cdots, N$, $\mu, \nu = 0, \cdots, n - 1$) 
must be (equivalent to) the sum of $N$ separate Pauli-Fierz actions, 
namely\footnote{We use
the signature ``mostly plus'': $- + + + \cdots$.
Furthermore, spacetime
indices are raised and lowered with the flat Minkowskian metric
$\eta_{\mu \nu}$. Finally, we take the spacetime dimension $n$ to be
strictly greater than $2$ since otherwise, the Lagrangian is
a total derivative. Gravity in two dimensions 
needs a separate treatment.} 
\beq 
\label{startingpoint} 
S_0[h_{\mu \nu}^a] &=& \sum_{a = 1}^N \int d^n x \ll[ 
-\frac{1}{2}\ll(\pa_{\m}{h^a}_{\n\r}\rr)\ll(\pa^{\m}{h^a}^{\n\r}\rr) 
+\ll(\pa_{\m}{h^a}^{\m}_{~\n}\rr)\ll(\pa_{\r}{h^a}^{\r\n}\rr)\rr.\nn\\ 
&&\ll.-\ll(\pa_{\n}{h^a}^{\m}_{~\m}\rr)\ll(\pa_{\r}{h^a}^{\r\n}\rr) 
+\frac{1}{2}\ll(\pa_{\m}{h^a}^{\n}_{~\n}\rr) 
\ll(\pa^{\m}{h^a}^{\r}_{~\r}\rr)\rr] \, , \; \; n>2. 
\eeq 
As we shall see below, our treatment, which is purely algebraic, 
extends (at least formally) to the case where the the collection
$\{h_{\mu \nu}^a\}$ is, possibly uncountably, infinite. However, for
simplicity, we  consider in most of the text a finite collection of
$N$ massless spin-two fields.

The action (\ref{startingpoint})
 is invariant under the following linear gauge transformations, 
\beq 
\delta_\epsilon h^a_{\mu \nu} = \partial_\mu \epsilon_\nu^a 
+ \partial_\nu \epsilon_\mu^a 
\label{freegauge} 
\eeq 
where the $\epsilon_\nu^a$ are $n \times N$ arbitrary, independent functions.
These transformations are abelian and irreducible.  
 
The equations of motion are 
\be 
 \frac{\d S_0}{\d h^a_{\mu \nu}} \equiv - 2 H_a^{\mu \nu} = 0 
\ee 
where $H^a_{\mu \nu}$ is 
the linearized Einstein tensor,
\beq 
H^a_{\mu \nu} =  K^a_{\mu \nu} - \frac{1}{2} K^a \eta_{\mu \nu}. 
\eeq 
Here, $K^a_{\a \b \m \n}$  is the linearized Riemann tensor, 
\be 
K^{a}_{\a\b\m\n}=- \frac{1}{2}
(\pa_{\a\m}h_{\b\n}^{a}+\pa_{\b\n}h_{\a\m}^{a} 
-\pa_{\a\n}h_{\b\m}^{a} 
-\pa_{\b\m}h_{\a\n}^{a})\,, 
\ee 
$K^a_{\mu \nu}$ is the linearized Ricci tensor, 
\be 
K^a_{\mu \nu} = K^{a \a}_{\ . \,\m \a \n} = - \frac{1}{2}
(\Box h^a_{\mu \nu} + \cdots),
\ee 
and $K^a$ is the linearized scalar curvature, $K^a = \eta^{\mu \nu} 
K^a_{\mu \nu}$.  
The Noether identities expressing the invariance of the free 
action (\ref{startingpoint}) under (\ref{freegauge}) are 
\beq 
\partial_\nu H^{a \mu \nu} = 0 
\label{linBianchi} 
\eeq 
(linearized Bianchi identities).  The gauge symmetry removes unwanted 
unphysical states. 
 
The problem of introducing consistent interactions for a collection 
of massless spin-2 fields  
is that of adding local interaction terms 
to the action (\ref{startingpoint}) 
while modifying at the same time  
the original gauge symmetries if necessary, in such a 
way that the modified action be invariant under the modified gauge 
symmetries.  We shall exclusively consider interactions that
can formally be expanded in powers of a deformation parameter
$g$ (``coupling constant") and that are consistent order by  
order in $g$.  The class of ``consistent interactions" 
for (\ref{startingpoint}) studied
here could thus be called  more accurately ``perturbative,
gauge-consistent interactions" (since we focus on compatibility
with gauge-invariance order by order in $g$), but 
we shall just use the terminology ``consistent interactions"
for short.

Since we are interested in the 
classical theory, we shall also demand that the interactions contain at
most two derivatives\footnote{in the sense of the usual
power counting of perturbative field theory. 
Thus we allow only
terms that are quadratic in the first derivatives of $h^a_{\mu \nu}$
or linear in their second derivatives.}  
so that the nature of the differential 
equations for $h^a_{\mu \nu}$ is unchanged.
On the other hand, we shall make no assumption on the 
polynomial order of the fields in the Lagrangian or in the gauge 
symmetries. 

In an interesting work \cite{Wald1},  Cutler and Wald have 
proposed theories involving a multiplet of spin-2 fields, 
based on associative, commutative algebras.  These authors 
arrived at these structures by focusing 
on the possible structures of modified gauge transformations and their algebra.
However, they 
did not analyse the extra conditions that must be imposed 
on the modified gauge symmetries if these are to be compatible  
with a Lagrangian having the (unique, consistent) 
free field limit prescribed above. 
[Their work was subsequently extended to supergravity 
in \cite{Anco:1998mf}.] 
Some explicit examples of 
Lagrangians that realize the Cutler-Wald algebraic 
structures have been constructed 
in \cite{Wald2} and \cite{ovrut}, but none of these has 
an acceptable free field limit.  Indeed, their free field 
limit does involve a sum of Pauli-Fierz Lagrangians, but some of the 
``gravitons'' come with the wrong sign and thus, the energy 
of the theory is unbounded from below.  
To our knowledge, 
the question of whether other examples of (real) Lagrangians realizing 
the Cutler-Wald structure (with a finite number of gravitons) 
would exist and whether some of them 
would have a physically acceptable free field limit was left open. 
  
Motivated by these developments, we have re-analyzed the 
question of consistent interactions for a collection 
of massless spin-2 fields by imposing from the outset that the 
deformed Lagrangian should have the free field limit 
(\ref{startingpoint}). As we shall see,
it turns out that this requirement forces 
one additional condition on the Cutler-Wald algebra defining 
the interaction, namely, that it  
be ``symmetric'' with respect to the scalar product defined by the 
free Lagrangian (see below for the precise meaning of ``symmetric'').
This extra constraint is quite stringent and 
implies that the algebra is the direct sum of one-dimensional ideals. 
This eliminates all the cross-interactions between the various
gravitons\footnote{The extra 
condition is in fact also derived by different methods in 
\cite{Anco:1998mf}, (Eqs. (3.38) and (A.55)), but its 
full implications regarding the impossibility of cross-interactions 
have not been investigated.}. 
Let us state the main (no-go) result of this paper, spelling out
explicitly our assumptions :

\begin{theorem}
Under the assumptions of: locality, Poincar\'e invariance, 
Eq.(\ref{startingpoint}) as free field limit
and at most two derivatives in the Lagrangian, the only consistent 
deformation of Eq.(\ref{startingpoint}) 
involving a collection of spin-2 fields is 
(modulo field redefinitions) a sum of
independent Einstein-Hilbert (or possibly Pauli-Fierz) actions,
\be 
S[g^a_{\mu \nu}] = \sum_a \frac{2}{\kappa_a^2}\int d^nx 
(R^a  - 2 \Lambda^a) \sqrt{-g^a}, \; \; \; g^a_{\mu \nu} = 
\eta_{\mu \nu} +  \kappa^a h^a_{\mu \nu}  \, ,
\label{sum} 
\ee 
where $R^a$ is the scalar curvature of $g^a_{\mu \nu}$,
$g^a$ its determinant, $\kappa^a \geq 0$ a self-coupling 
constant and $\Lambda^a$ independent cosmological
constants. [A term with $\kappa^a = 0$ is a
Pauli-Fierz action; the corresponding cosmological term reads
$\lambda^a h^{a\m}_{\; \; \m}$.] In the case of an infinite
collection of spin-2 fields the sum in (\ref{sum}) may contain,
besides a series, an integral over continuous parameters.
\end{theorem}

There are  no other (perturbatively gauge-consistent)
possibilities under the assumptions stated.

We have also investigated how matter couplings  affect the problem
of the (non-~)existence of cross-interactions between gravitons.  
We have taken the simplest example of a scalar field and have 
verified that the scalar field can only couple to one type 
of gravitons. 
Thus, even the existence of indirect cross-couplings (via intermediate
interactions) between  massless spin-2 particle is 
excluded.
The interacting theory  
describes parallel worlds, and, in any given world, there is 
only one massless spin-2 field.  This massless spin-2 field has
(if it interacts at all) 
the standard graviton couplings with the fields living 
in its world (including itself), in agreement with 
the single massless spin-2 field studies of  
\cite{Gupta,Kraichnan,Feynman,Weinberg,Ogiev,Wyss,Deser1,Fronsdal,
Berendsetal,Wald0}. 

The above theorem relies strongly on the assumption that
the interaction contains at most two derivatives.
If one allows more derivatives in the Lagrangian,
one can construct 
cross-interactions involving the linearized curvatures,
which are manifestly consistent with gauge invariance.
An obvious cubic candidate
is
\be
g_{abc} K^a_{\a \b \m \n} K^{b \a \b}_{\rho \sigma}
K^{c \m \n \rho \sigma}
\label{foot}
\ee
where $g_{abc}$ are arbitrary constants.  This candidate
can be added to the free Lagrangian and defines an interacting
theory with the same abelian gauge symmetries as the original theory since
(\ref{foot}) is invariant under (\ref{freegauge}). It
contains six derivatives. 
Other deformations of the original free action 
that come to mind are obtained
by going to the Einstein theory
and adding then, in each sector,
higher order polynomials
in the curvatures and their covariant derivatives.

All these deformations have the important feature of deforming
the algebra of the gauge symmetries in a rather simple way:
the deformed algebra is the direct sum of independent diffeomorphism algebras
(in each sector with $\kappa_a \not=0$)
and abelian algebras.
This is not an
accident.  The possibilities of deformations of the
gauge algebra are in fact severely limited even in the
more general context where no constraint on the number of
derivatives is imposed (except that it should
remain bounded).  One has the theorem :

\begin{theorem}
Under the assumptions of locality, Poincar\'e invariance
and Eq.(\ref{startingpoint}) as free field limit,
the only consistent
deformations of  the action (\ref{startingpoint})
involving a collection of spin-2 fields are 
such that the algebra of the deformed gauge-symmetries  
is given, to first
order in the deformation parameter,
by the direct sum of independent
diffeomorphism algebras. 
[Some terms in the direct sum may remain undeformed, i.e.,
abelian.] In the case of an infinite collection, the direct
sum may include a continuous integral. 
\end{theorem}        

This theorem strengthens previous results
in that it does not assume off-shell closure
of the gauge algebra (this is automatic) or any specific form
of the gauge symmetries (which are taken to involve only one
derivative in most treatments).

In order to prove these results, 
we shall begin the analysis
without making any assumption on the number
of derivatives, except that it is bounded.  
We shall see that this is indeed enough to completely control
the algebra.  
We shall then point out where the derivative assumptions
are explicitly needed, at the level of the gauge
transformations and of the deformation of the Lagrangian.
We shall discuss in section
\ref{derivatives} the new features that appear in
the absence of these assumptions.

Our approach is based on the BRST reformulation of the problem, in 
which consistent couplings define deformations of the 
solution of the so-called ``master equation''.   
The advantage of this approach 
is that it clearly organizes the calculation of the non-trivial consistent 
couplings in terms of cohomologies which are either already known
 or easily computed. 
These cohomologies are in fact interesting in themselves, besides 
their occurence in the consistent interaction problem.  One of them is 
the ``characteristic cohomology'', which investigates higher order 
conservation laws involving antisymmetric tensors (see below).  
The use of BRST techniques 
somewhat streamlines the derivation, which would otherwise 
be more cumbersome. 
 
In the next section, we review the master-equation approach to the 
problem of consistent interactions.  We then recall some  
cohomological results necessary for solving the problem.  In 
particular, we discuss at length the characteristic cohomology 
(section \ref{characteristic}).  Section \ref{hard} constitutes the 
 core of our paper.  We show how the structure of an associative, 
commutative algebra introduced first in this 
context by Cutler and Wald arises in the cohomological approach, 
and derive the further crucial condition of ``symmetry" (explained 
in the text) that emerges from the requirement that the deformation  
not only defines consistent gauge transformations, but also can be 
extended to a consistent deformation of the Lagrangian.  We then 
show that all the requirements on the algebra force it to be trivial 
(section \ref{obstr}), which implies that there can be no 
cross-interaction between the various spin-2 fields. In the 
next section (section \ref{complete}),  
we complete the construction of the consistent 
Lagrangians and establish the validity of (\ref{sum}). 
In section \ref{infinitedim}, we comment on the
case with an infinite number of different types of
gravitons.

Section \ref{mattercoupling} shows that the coupling to matter 
does not allow the different types of gravitons to ``see each 
other'' through the matter.  In section \ref{nondefinitepositive} 
we  briefly generalize the 
discussion to the presumably
physically unacceptable case of
non-positive metrics in the internal space of the 
gravitons. This is done solely for the sake of comparison
with the work of \cite{Wald1,Wald2},   where there are
propagating ghosts. 
Section \ref{derivatives} discusses the 
new features that arise when no restriction is imposed 
on the number of derivatives in the Lagrangian. 
A brief concluding section is finally followed by 
a technical appendix that collects the proofs of the theorems 
used in the core of the paper.

\section{Cohomological reformulation} 
\setcounter{equation}{0} 
\setcounter{theorem}{0} 
\setcounter{lemma}{0} 
 
\subsection{Gauge symmetries and master equation} 
The central idea behind the master equation approach to the 
problem of consistent deformations is the following.  Consider 
an arbitrary irreducible gauge theory  
with fields $\Phi^i$, 
action $S[\Phi^i]$, gauge transformations\footnote{Throughout this
section, we use De Witt's condensed 
notation  in which a summation over a repeated index 
implies also an integration. The $R^i_{\a}\ll(\Phi\rr)$ stand for 
$R^i_{\a}(x,x')$ and are combinations of the Dirac 
delta function $\delta(x,x')$ and some of its derivatives with coefficients 
that involve the fields and their derivatives, so that 
$R^i_{\a} \e^{\a} \equiv \int d^n x' R^i_{\a}(x,x') \e^{\a}(x')$
is a sum of integrals of $\e^{\a}$ and a finite number of its derivatives.} 
\be 
\d_{\e}\Phi^i=R^i_{\a}\ll(\Phi\rr)\e^{\a}\,, 
\ee 
and gauge algebra 
\be 
R^j_{\a}\ll(\Phi\rr)\frac{\d R^i_{\b}\ll(\Phi\rr)}{\d \Phi^j} 
-R^j_{\b}\ll(\Phi\rr)\frac{\d R^i_{\a}\ll(\Phi\rr)}{\d \Phi^j} 
=C^{\gamma}_{\a\b}\ll(\Phi\rr)R^i_{\gamma}\ll(\Phi\rr)\, 
+ M_{\a\b}^{i j}\ll(\Phi\rr) \frac{\d S}{\d \Phi^j}. 
\label{gaugealgebra} 
\ee 
We have allowed the gauge transformations to close only 
on-shell. The coefficient functions $M_{\a\b}^{i j}$ are (graded) 
antisymmetric 
in both $\a$, $\b$ and $i$, $j$.  The Noether identities read 
\be 
\frac{\d S}{\d \Phi^i} R^i_{\a} = 0. 
\label{Noether} 
\ee 
One can derive higher order 
identities from (\ref{gaugealgebra}) and (\ref{Noether}) 
by differentiating (\ref{gaugealgebra}) with respect to the fields  
and using the fact that second-order derivatives 
commute.  These identities, in turn, lead to further identities 
by a similar process. 
 
It has been established in \cite{dwvh,bv1} that one 
can associate with $S$ a functional $W$  depending
on the original fields $\Phi^i$ and on additional variables, called the 
ghosts $C^\a$ and the antifields $\Phi^*_i$ and $C^*_\a$,
with the following properties: 
\begin{enumerate} 
\item $W$ starts like 
\be 
\label{genericmaster} 
W = S + \Phi^*_iR^i_{\a}C^{\a}+\frac{1}{2} 
C^*_{\gamma}C^{\gamma}_{\a\b}C^{\b}C^{\a} 
+ \frac{1}{2} \Phi^*_i \Phi^*_j M_{\a\b}^{i j} C^\a C^\b  
+ \hbox{ ``more''} 
\ee 
where ``more'' contains at least three ghosts; 
\item $W$ fulfills the equation 
\be 
(W,W) = 0 
\label{mastereq} 
\ee 
in the antibracket $(,)$ that makes the fields and the antifields 
canonically conjugate to each other.  This antibracket structure 
was first introduced by Zinn-Justin\footnote{In Zinn-Justin's work 
the antifields appear as ``sources'' $K_i, L_\a$.}\cite{ZJ}
 and was denoted originally by a  
$\star$ ($(A,B) \equiv A \star B$).  It is defined by 
\be 
(A,B) =  \frac{\d^R A}{\d \Phi^i} \frac{\d^L B}{\d \Phi^*_i} 
- \frac{\d^R A} {\d \Phi^*_i} \frac{\d^L B}{\d \Phi^i} 
+ \frac{\d^R A}{\d C^\a} \frac{\d^L B}{\d C^*_\a} 
- \frac{\d^R A}{\d C^*_\a} \frac{\d^L B}{\d C^\a}, 
\ee 
where the superscript $R$ ( $L$) denotes a right ( left) derivative,
respectively.
 
\item $W$ is bosonic and has ghost number zero.
\end{enumerate} 
To explain this last statement, we
recall that all fields belong to a Grassmann algebra $\cal G$: the fields
$\Phi^i$ and $C^*_\a$ belong to the even part of $\cal G$ (i.e.
they commute with everything), while the fields $C^\a$ and $\Phi^*_i$
belong to the odd part of $\cal G$ (i.e. they anticommute among
themselves). [Instead of ``commuting'' or ``anticommuting'', we shall
simply say ``bosonic'', or ``fermionic'', respectively. Note, however,
that we work in a purely classical framework.]
Moreover, in addition to the above ``fermionic''
$Z_2$ grading
(odd or even) one endows the algebra of the dynamical variables
with a Z-valued
``ghost grading'' defined such that the original
fields $\Phi^i$, the ghosts
$C^{\a}$, the antifields $\Phi^*_i$ and the antifields $\Phi^*_{\a}$
have ghost number zero, one, minus one and minus two, respectively.           
The statement that $W$ has ghost number zero means 
that each term in $W$ has a zero ghost number.
Note that the antibracket increases the ghost number by
one unit, i.e., $gh((A,B)) = gh(A) + gh(B) + 1$ (we refer
to the book \cite{BOOK} for more information).             

It is also useful to introduce a second Z-valued grading 
for the basic variables, called
the ``antifield'' (or ``antighost'') number \cite{BOOK}. 
This grading is defined by assigning antifield number
zero to the fields $\Phi^i$ and the ghosts $C^\a$, antifield
number one to the antifields $\Phi^*_i$ and antifield number
two to the antifields $C^*_\a$.  The antifield number thus counts 
the number of antifields $\Phi^*_i$ and $C^*_\a$,
with weight two given to the antifields $C^*_\a$ conjugate
to the ghosts.
There are different ways to achieve a fixed ghost number
by combining the antifields and the ghosts. For instance,
$\Phi^*_i C^\a$, $C^*_\a C^b C^c$, $\Phi^*_i C^*_\a C^b C^c C^d$
all have ghost number zero; but the first term has antifield
number one, the second has antifield number two and the third has
antifield number three.  The antifield number keeps track
of these different possibilities.  By introducing it, one
can split an equation with definite ghost number into simpler
equations at each value of the antifield number.  This procedure
will be amply illustrated in the sequel.

In our irreducible case where there is only one type of ghosts,
the antifield number can also be viewed as 
an indirect way of keeping track of the number of explicit
ghost fields $C^a_\a$ entering any expression. Indeed, if we define the
``pureghost number'' of any expression as the number of explicit $C^a_\a$'s
in it, it is easy to see from the antighost attributions above
that the (net) ghost number is given by: $gh = puregh - antigh$.  

The equation (\ref{mastereq}) is called the master equation 
while the function $W$ is called  
the (minimal) solution of the master equation. 
It is easily seen that, because of the $Z_2$-grading of
the various fields (the ``canonically conjugate'' fields in the antibracket
have opposite fermionic gradings), $(A,B)$ is {\it symmetric} for
bosonic functions $A$ and $B$, $(A,B)= (B,A)$. One can also check that
the antibracket satisfies the (graded) Jacobi identity (see, e.g., \cite{BOOK}).
This fact will play an important role in the work below. 
 
The master equation is fulfilled as a consequence of 
the Noether identities (\ref{Noether}), of the gauge 
algebra (\ref{gaugealgebra}) and of all the higher order 
identities alluded to above that one can derive from them. 
Conversely, given some $W$, solution of (\ref{mastereq}), one can recover 
the gauge-invariant action as the term independent of the ghosts 
in $W$, while the gauge transformations are defined by the terms 
linear in the antifields $\Phi^*_i$ and the structure functions 
appearing in the gauge algebra can be read off from the terms 
quadratic in the ghosts.  The Noether identities  
(\ref{Noether}) are fulfilled 
as a consequence of the master equation (the left-hand side of the 
Noether identities is the term linear in the ghosts in $(W,W)$; 
the gauge algebra (\ref{gaugealgebra}) is the next term in 
$(W,W) = 0$). 
 
In other words, there is complete equivalence between gauge invariance 
of $S$ and the existence 
of a solution $W$ of the master equation. 
For this reason, one can reformulate the problem of consistently 
introducing interactions for a gauge theory as that of 
deforming $W$ while maintaining the master equation \cite{bh}. 
 
\subsection{Perturbation of the master equation} 
Let $W_0$ be the solution of the master equation for the 
original free theory,  
\be 
W_0 = S_0 + \Phi^*_iR^i_{0\a}C^{\a},  \; (W_0,W_0) = 0. 
\ee 
Because the gauge transformations are abelian, there is no 
further term in $W_0$ ( $C^{\gamma}_{\a\b} =0, 
M_{\a\b}^{i j} =0 $). 
Let $W$ be the solution of the master 
equation for the searched-for interacting theory, 
\begin{eqnarray} 
W &=& S + \Phi^*_iR^i_{\a}C^{\a} + O(C^2), \\ 
S &=& S_0 + \hbox{interactions}, \\ 
R^i_{\a} &=& R^i_{0\a} + \hbox{deformation terms} ,\\ 
(W,W) &=& 0. 
\end{eqnarray} 
As we have just argued, $W$ exists if and only if 
$S = S_0 + \hbox{``interactions''}$ 
is a consistent deformation of $S_0$. 

Let us now expand $W$ and the master equation for $W$ in powers of the
deformation parameter $g$.  With
\be 
W = W_0 + g W_1 + g^2 W_2 + O(g^3) 
\ee 
the equation $(W,W)= 0$ yields, up to order $g^2$ 
\begin{eqnarray} 
O(g^0): & & (W_0,W_0) = 0 \label{key1}\\ 
O(g^1): & & (W_0,W_1) = 0 \label{key2}\\ 
O(g^2): & & (W_0,W_2) = - \frac{1}{2} (W_1,W_1). \label{key3} 
\end{eqnarray} 
The first equation is fulfilled by assumption since the starting point 
defines a consistent theory.  To analyse the higher order equations, 
one needs further information about the meaning of $W_0$. 
 
\subsection{BRST transformation, first order deformations, obstructions} 
It turns out that $W_0$ is in fact the generator of the BRST transformation 
$s$ of the free theory through the antibracket\footnote{We denote 
the BRST transformation for the free theory by $s$, rather 
than $s_0$ because this is the only BRST symmetry we shall consider 
so no confusion can arise.}, i.e. 
\be 
sA = (W_0, A). 
\label{BRSTcocy} 
\ee 
The nilpotency $s^2 = 0$ follows from the master equation 
(\ref{key1}) for $W_0$ and the (graded) Jacobi identity 
for the antibracket.  Thus, Eq. (\ref{key2}) simply expresses 
that $W_1$ is a BRST-cocycle, i.e. that it is ``closed'' under $s$:
$s W_1 = 0$.  
 
Now, not all consistent interactions are relevant.  Indeed, one 
may generate ``fake'' interactions by making non-linear field 
redefinitions.  Such interactions are trivial classically and 
quantum-mechanically \cite{Tyutin:2000ht}.  One 
can show \cite{bh} that the physically trivial interactions 
generated by field-redefinitions that reduce to the identity at order $g^0$, 
\be 
\Phi^i \rightarrow \Phi'^i = \Phi^i + g \, \Xi^i(\Phi, \partial \Phi, \cdots) 
+ O(g^2) 
\ee                               
precisely correspond to cohomologically trivial solutions 
of (\ref{BRSTcocy}), i.e.,correspond to ``exact'' $A$'s (also
called ``coboundaries'') of the form 
\be 
A = s B 
\label{BRSTcoboun} 
\ee 
for some $B$.  We thus come  to the conclusion that the non-trivial  
consistent interactions are characterized to first 
order in $g$ by the {\it cohomological group}\footnote{ We recall that,
given some nilpotent $s$, $s^2 =0$, $H(s)$ denotes the equivalence
classes of ``closed'' $A$'s, modulo ``exact'' ones, i.e. the
solutions of $ s A =0$, modulo the equivalence relation 
$ A' = A + s B$.} $H(s)$ at ghost number zero. 
In fact, since $W_1$ must be a local functional, the cohomology 
of $s$ must be computed in the space of local functionals. 
Because the equation $s \int a = 0$ is equivalent to $sa + dm = 0$ 
(where $d$ denotes Cartan's exterior differential)
for some $m$, and $\int a = s \int b$ is equivalent to $a = s b + dn$ 
for some $n$, one denotes the corresponding cohomological 
group by $H^{0,n}(s \vert d)$\footnote{More generally, we shall use in this
paper the notation $H^{i,p}_{j}$ to denote a cohomological group for 
$p-$forms having a fixed ghost number $i$, and a fixed ``antifield''
number $j$ (see below). If we indicate only one superscript, it will
always refer to the form degree $p$.} ($0$ is the ghost number and $n$ the  
form-degree: $a$ and $b$ are $n$-forms).  
 
The redundancy in $W_1$ is actually slightly bigger than the possibility 
of adding trivial cocycles, since one can admit changes of field 
variables $\Phi^i \rightarrow \Phi'^i$ that do not 
reduce to the identity at zeroth order in $g$, but reduce to a global symmetry 
of the original theory, i.e., leave 
the free action invariant. 
Two distinct BRST cocycles $W_0$ 
and $W'_0$ that can be 
obtained from one another under such a transformation should be identified. 
In practice, however, only a few of these transformations are to be 
taken into account (if any) since only a few of them preserve the condition 
on the number of derivatives of the deformation.   
This will be explicitly illustrated in the graviton case. 
 
Once a first-order deformation is given, one must investigate 
whether it can be extended to higher orders. 
It is a direct consequence of the Jacobi 
identity for the antibracket that $(W_1,W_1)$ is BRST-closed, 
$(W_0,(W_1,W_1)) = 0$. 
However, it may not be BRST-exact (in the space of local functionals). 
In that case, the first-order deformation $W_1$ is obstructed at 
second-order, so, it is not a good starting point.  If, on 
the other hand, $(W_1,W_1)$ is BRST-exact, then a 
solution $W_2$ to (\ref{key3}), which may 
be rewritten 
\be  
sW_2 = -\frac{1}{2} (W_1,W_1), 
\ee 
exists. As $(W_1,W_1)$ has ghost number one (since the antibracket
 increases the ghost number by one unit), we see  that obstructions 
 to continuing a given, first-order consistent 
interaction are  measured by the cohomological group 
$H^{1,n}(s \vert d)$.  Furthermore, the ambiguity in $W_2$ 
(when it exists) is a solution of the homogeneous equation 
$(W_0,W_2) = 0$.  Among these solutions, 
those that are equivalent through  field redefinitions should 
be identified. $O(g^2)$-redefinitions of 
the fields  yield trivial BRST-cocycles, so again, the space of 
equivalent  
$W_2$'s is a quotient of $H^{0,n}(s \vert d)$. 
Further identifications follow from $O(g^0)$ and $O(g^1)$-redefinitions 
that leave the previous terms invariant.  These identifications 
will be discussed in more details below. 
 
The same pattern is found at higher orders : obstructions  
to the existence of $W_k$ are 
elements of $H^{1,n}(s \vert d)$, while the ambiguities  
in $W_k$ (when it exists) are 
elements of appropriate quotient spaces of $H^{0,n}(s \vert d)$.   
 
Since the identifications of equivalent solutions will play an important 
role in the sequel, let us be more explicit on the precise form 
that the equations describing these identifications take. 
Two solutions of the master 
equation are equivalent if they differ by an anti-canonical transformation 
in the  
antibracket.  These correspond indeed to field and gauge 
parameter (ghost) redefinitions 
\cite{Dixon,Voronov,BOOK}. 
Infinitesimally, two solutions $W$ and $W + \Delta W$ are 
thus equivalent if 
\be 
\Delta W = (W, K) 
\ee 
for some $K$ of ghost number $-1$. 
If we expand this equation in powers of $g$, we get 
\beq 
\Delta W_0 &=& (W_0,K_0), \label{deltaw0} \\ 
\Delta W_1 &=& (W_0,K_1) + (W_1,K_0), \label{deltaw1}\\ 
\Delta W_2 &=& (W_0,K_2) + (W_1,K_1)+ (W_2,K_0),  \label{deltaw2}\\ 
\cdots && \nonumber 
\eeq 
Since $W_0$ is given, one must impose $\Delta W_0 = 0$, and thus, 
from (\ref{deltaw0}),  
\be 
(W_0,K_0) = 0: \label{globalsymmetry} 
\ee 
$K_0$ defines a global symmetry of the free theory \cite{BBH1,BBH5}. 
The first term on the right-hand side of (\ref{deltaw1}) is 
a BRST-coboundary and shows that indeed, one must identify 
two BRST-cocycles that are in the same cohomological class 
of $H^{0,n}(s \vert d)$.  There is a further identification  
implied by the term $(W_1,K_0)$.   Similarly, besides the 
BRST-coboundary $(W_0,K_2)$, there are extra terms in the right-hand side 
of (\ref{deltaw2}). 
 
The cohomological considerations that we have just outlined are 
equivalent to the conditions for consistent interactions 
derived in \cite{Berends} without use of ghosts or antifields. 
The interest of the master equation approach is that it organizes 
these equations in a rather neat way.  Also, one can use  
cohomological tools, available in the literature,
 to determine these interactions and their obstructions. 
 
In the sequel, we shall compute explicitly $H^{0,n}(s \vert d)$ for 
a collection of free, massless spin-2 fields, i.e., we shall determine 
all possible first-order consistent interactions.  We shall then determine 
the conditions that these must fulfill in order to 
be unobstructed at order $g^2$.  These 
conditions turn out to be extremely strong and prevent 
cross interactions between the various types of gravitons. 
 
\subsection{Solution of the master equation for a collection of free, 
spin-2, massless fields} 
 
We rewrite the free action (\ref{startingpoint}) as 
\beq 
\label{PFalgebradelta} 
S_0&=&\int d^nx~k_{ab}\ll[ 
-\frac{1}{2}\ll(\pa_{\m}{h^a}_{\n\r}\rr)\ll(\pa^{\m}h^{b\n\r }\rr) 
+\ll(\pa_{\m}h^{a \m}_{~\n}\rr)\ll(\pa_{\r}h^{b \r\n}\rr)\rr.\nn\\ 
&&\ll.-\ll(\pa_{\n}h^{a \m}_{~\m}\rr)\ll(\pa_{\r}h^{b \r\n}\rr) 
+\frac{1}{2}\ll(\pa_{\m}h^{a \n}_{~\n}\rr) 
\ll(\pa^{\m}h^{b \r}_{~\r}\rr)\rr]\,, 
\eeq
with a quadratic form $ k_{a b}$ defined by the kinetic terms. In the 
way of writing the Pauli-Fierz free limit above, Eq.(\ref{startingpoint}),
$k_{ab}$ was simply the Kronecker delta $\d_{ab}$. What is
essential for the physical consistency of the theory (absence of
negative-energy excitations, or stability of the Minkowski vacuum) is that
$ k_{a b}$ defines a positive-definite metric 
in internal space; it can then be normalized to be  $\delta_{ab}$ by
a simple linear field redefinition. 
 
Following the previous prescriptions, the fields, ghosts 
and antifields are found to be 
\begin{itemize} 
\item the fields $h^a_{\a\b}$, with ghost number zero and antifield number zero; 
\item the ghosts $C^a_{\a}$, with ghost number one and antifield number zero; 
\item the antifields $h^{* \a\b}_{a}$, with ghost  
number minus one and antifield  
number one; 
\item the antifields $C^{* \a}_{a}$, with ghost number minus  
two and antifield  
number two. 
\end{itemize} 
 
The solution of the master equation for the free theory is, reverting 
to notations where integrals are all explicitly written, 
\be 
W_0 = S_0 + \int d^nx \, h^{* \a\b}_{a} (\partial_\a C 
+ \partial_\b C^a_\a), 
\ee  
from which we get the BRST differential $s$ of the free 
theory as  
\be 
s = \delta + \gamma 
\ee 
where the action of $\gamma$ and $\delta$ on the variables is  
zero except 
(note in particular that $ \g C^a_\a = \d C^a_\a =0$)\footnote{
We denote $t_{(\a\b)} \equiv \frac{1}{2} (t_{\a\b} + t_{\b\a})$, and
$t_{[\a\b]} \equiv \frac{1}{2} (t_{\a\b} - t_{\b\a})$.}
\beq 
\g h^a_{\a\b}&=&2\pa_{(\a}C^a_{\b)} \label{trivialdc} \label{defg}\\ 
\d h_a^{*\a\b}&=&\frac{\d S_0}{\d h^a_{\a\b}}\label{defd1}\\ 
\d C_a^{*\a}&=&-2\pa_{\b}h_a^{*\b\a} . \label{defd2} 
\eeq 
The decomposition of $s$ into $\delta$ plus $\gamma$ is 
dictated by the antifield number: $\delta$ decreases the 
antifield number by one unit, while $\gamma$ leaves it unchanged. 
Combining this property with $s^2 =0$, one concludes that, 
\be 
\delta^2 = 0, \; \delta \gamma + \gamma \delta = 0, \; 
\gamma^2 = 0. 
\ee

\section{Cohomology of $\gamma$} 
\setcounter{equation}{0} 
\setcounter{theorem}{0} 
\setcounter{lemma}{0} 
\label{cohoofg} 
 
To compute the consistent, first order deformations, i.e., 
$H(s \vert d)$, we shall see in Section 5 that we need 
$H(\gamma)$ and $H(\delta \vert d)$. 
We start with $H(\gamma)$, which is rather easy. 
 
As it is clear from its definition, $\gamma$ is related to the 
gauge transformations.  Acting on anything, it gives zero, except 
when it acts on the spin-2 fields, on which it gives a gauge transformation 
with gauge parameters replaced by the ghosts. 
 
The only gauge-invariant objects that one can construct out of 
the gauge fields $h^a_{\mu \nu}$ and their derivatives are the 
linearized curvatures $K^a_{\a \b \m \n}$ and their derivatives. 
 
The antifields and their derivatives are also $\gamma$-closed.  
The ghosts and their derivatives are $\gamma$-closed as well but 
their symmetrized first order derivatives are $\gamma$-exact 
(see Eq. (\ref{trivialdc})), as 
are all their subsequent derivatives since 
\be 
\label{secondder} 
\pa_{\a\b}C^a_{\g}=\frac{1}{2}\,\g\ll(\pa_{\a}h^a_{\b\g}+\pa_{\b}h^a_{\a\g} 
-\pa_{\g}h^a_{\a\b}\rr)\,. 
\ee 
 
It follows straightforwardly from these observations 
that the $\g$-cohomology is generated 
by the linearized curvatures, the antifields and all their derivatives, as 
well as by the 
ghosts $C^a_\mu$ and their antisymmetrized first-order derivatives 
$\pa_{[\m}C^a_{\n]}$.  More precisely, let $\{\omega^I\}$ be a basis 
of the space of polynomials in the $C^a_\mu$ and $\pa_{[\m}C^a_{\n]}$ (since 
these variables anticommute, this space is finite-dimensional).  One has: 
\be 
\label{alphaomega} 
\g a = 0 \Rightarrow a =  
\a_J\ll([K],[h^*],[C^*]\rr)\omega^J\ll(C^a_{\m},\pa_{[\m}C^a_{\n]}\rr) 
+ \gamma b \,, 
\ee 
where the notation $f([m])$ means that the function $f$ depends 
on the variable $m$ and its subsequent derivatives up to a finite order. 
If $a$ has a fixed, finite ghost number, then $a$ can 
only contain a finite number of antifields.  If we assume in addition 
that $a$ has a bounded number of derivatives, as we shall do from now on, 
then, the $\a_J$ are polynomials\footnote{A term like $\exp \kappa K$, where 
$K$ is the linearized scalar curvature, does not have a 
bounded number of derivatives since it contains  
arbitrarily high powers of $K$, and since the number of derivatives in $K^m$ 
is $2m$.  Note, however, that the coefficient 
at each order in the coupling $\kappa$ is of bounded derivative 
order - just by dimensional analysis -  so that in our perturbative approach 
where we expand the interactions in powers of the coupling constant 
and work order by order, the assumption of bounded derivative 
order is not a restriction.}.   
 
In the sequel, the polynomials $\a_J\ll([K],[h^*],[C^*]\rr)$ in the 
linearized curvature $K^a_{\a \b \m \n}$, the antifields $h^{* \m \n}_a$ 
and $C^{* \m}_a$, as well as all their derivatives, will be 
called ``invariant polynomials''.  They may of course  
have an extra, unwritten, 
dependence on $dx^\m$, i.e., be exterior forms. 
At zero antifield number, the invariant polynomials are 
the polynomials in the linearized curvature $K^a_{\a \b \m \n}$ 
and its derivatives. 
 
We shall need the following theorem on the cohomology of $d$ in the 
space of invariant polynomials. 
\begin{theorem}\label{2.2} 
In form degree less than $n$ and in 
antifield number strictly greater than $0$, 
the cohomology of $d$  
is trivial in the space of invariant 
polynomials. 
\end{theorem} 
That is to say, if $\a$ is an invariant polynomial, the equation 
$d \a = 0$ with $antigh(\a) > 0$ implies   
$ \a = d \b$ where $\b$ is also an invariant polynomial. 
To see this, treat the antifields as ``foreground fields"
and the curvatures as ``background fields'', as in \cite{Duboisetal}.
Namely, split $d$ as $d = d_1 + d_0$, where $d_1$ acts only
on the antifields and $d_0$ acts only on the curvatures. The
so-called   ``algebraic
Poincar\'e lemma'' states that $d_1$ has no cohomology
in form degree less than $n$ (and in
antifield number strictly greater than $0$)
because there is no relation among the derivatives
of the antifields.  By contrast, $d_0$ has some cohomology in
the space of polynomials in the curvatures because these
are constrained by the Bianchi identities.  From the
triviality of the cohomology of $d_1$,
one easily gets $d \a = 0 \Rightarrow \a = d \b + u$, where $\b$ is
an invariant polynomial, and where $u$ is an invariant polynomial
that does not involve the antifields.  However, since $antigh(\a) > 0$,
$u$ must vanish. qed.

\section{Characteristic cohomology  -- cohomology of $\delta$ 
modulo $d$} 
\label{characteristic} 
\setcounter{equation}{0} 
\setcounter{theorem}{0} 
\setcounter{lemma}{0} 
 
\subsection{Characteristic cohomology} 

It has been shown in \cite{locality} that $H(\delta \vert d)$ 
is trivial in the space of forms with positive pure ghost number. Thus 
the next cohomology that we shall need is $H(\delta \vert d)$ in the space 
of local forms that do not involve the ghosts, i.e., having $puregh =0$.
This cohomology has an interesting interpretation in terms of conservation  
laws, which we first review \cite{BBH1} (see also \cite{BBH5} for 
a recent review). 
 
Conserved currents $j^\m$ are defined through the condition 
\be 
\partial_\m j^\m \approx 0 
\label{cocchar} 
\ee 
where $\approx$ means ``equal when the equations of motion hold'', 
or, as one also says, ``weakly equal to''.  These currents may carry 
further internal or spacetime indices that we shall not write explicitly. 
Among the conserved currents, those of the form 
\be 
j^\m_{\rm triv} \approx \partial_\n S^{\m \n} 
\label{cobchar} 
\ee 
where $S^{\m \n}$ is antisymmetric in $\m$ and $\n$, $S^{\m \n} 
= - S^{\n \m}$, are sometimes called (mathematically) trivial (although they 
may not be physically trivial), because they can be 
constructed with no information on 
the equations of motion. We shall adopt this terminology here. 
If we call $k$ the ($n-1$)-form dual to 
$j^\m$, and $r$ the ($n-2$)-form dual to $S^{\m \n}$, 
the conditions (\ref{cocchar}) and (\ref{cobchar}) 
can be rewritten as 
\be 
dk \approx 0 
\label{cocchar2} 
\ee 
and 
\be 
k_{\rm triv} \approx dr, 
\label{cobchar2} 
\ee 
respectively. 
These conditions define the characteristic cohomology $H^{n-1}_{\rm char}(d)$ 
in degree $n-1$ \cite{Vino,Bryant}. 
One may define more generally the characteristic cohomology 
$H^p_{\rm char}(d)$
in any form degree $ p \leq n$, by the same conditions  
(\ref{cocchar2}) and (\ref{cobchar2}). 
Again, $k$ may have extra internal or spacetime unspecified indices. 
 
\subsection{Cohomology of $\delta$ modulo $d$}

A crucial aspect of the differential $\delta$ defined through 
(\ref{defd1})  and (\ref{defd2}) is that it is related 
to the dynamics of the theory.  This is obvious since 
$\delta h^{*\ m \n}_a$ reproduces the Euler-Lagrange derivatives 
of the Lagrangian.  In fact, one has the following important 
(and rather direct) 
results about the cohomology of $\delta$ \cite{FH,locality,BOOK} 
\begin{enumerate} 
\item Any form of zero antifield number 
which is zero on-shell is $\delta$-exact; 
\item $H^p_i (\delta) = 0$ for $i>0$, where $i$ is the antifield 
number, in any form-degree $p$.   
[The antifield number is written as a lower index; the ghost number is not 
written because it is irrelevant here.] 
\end{enumerate} 
 
Because of the first property, one can rewrite the cocycle condition 
and coboundary condition of the characteristic cohomology as 
\be 
dk^p_0 + \delta k^{p+1}_1 = 0 
\label{cocchar5} 
\ee 
and 
\be 
k^p_{{\rm triv} 0} = d r^{p-1}_0 + \delta r^{p}_1, 
\ee 
respectively, 
where all relevant degrees have been explicitly written (recall
that there is no ghost here, i.e., 
$puregh = 0$ throughout section \ref{characteristic}).
Thus, we see that the characteristic cohomology is just 
$H_0^p(d \vert \delta)$.  Using $H^p_i (\delta) = 0$ for $i>0$, 
one can then easily establish the isomorphisms $H_0^p(d \vert \delta) 
\simeq H^{p+1}_1 (\delta \vert d)$ ($n> p > 0$ ) and $H_0^0(d \vert \delta) 
/ R \simeq H^1_1 (\delta \vert d)$ \cite{Duboisetal,BBH1}\footnote{The
quotient $H_0^0(d \vert \delta)/R$ is taken here in the sense of
vector spaces: the set $R$ of real numbers is naturally identified
with a vector
subspace of $H_0^0(d \vert \delta)$ since $dc = 0$ for
any constant $c$ and $c \not= \d$(something) $+ 
d$(something') unless $c = 0$. The constants
occur in the isomorphism $H_0^0(d \vert \delta)
/ R \simeq H^1_1 (\delta \vert d)$ because the cohomology
of $d$ is non trivial in form degree zero, $H^0(d) \simeq R$
(see \cite{Duboisetal} for details).  The relation
$H_0^0(d \vert \delta)
/ R \simeq H^1_1 (\delta \vert d)$ implies in particular that 
if $H^1_1 (\delta \vert d) = 0$, then $H_0^0(d \vert \delta) \simeq R$.}

Finally, using the isomorphism $H^i_j(\delta \vert d) \simeq 
H^{i+1}_{j+1}(\delta \vert d)$ \cite{BBH1}, we conclude 
\beq 
H^{n-p}_{\rm char}(d) &\simeq& H^n_p(\delta \vert d), \, 0 <p < n 
\label{redchar0}\\ 
H^0_{\rm char}(d)/R &\simeq& H^n_n (\delta \vert d) . 
\label{redchar}
\eeq 

The following vanishing theorem on $H^n_p(\delta \vert d)$
(and thus also on $H^{n-p}_{\rm char}(d)$ or 
$H^0_{\rm char}(d)/R$) can be proven: 
\begin{theorem} 
\label{vanishing} 
The cohomology groups $H^n_p(\delta \vert d)$ vanish 
in antifield number strictly greater than $2$, 
\be 
H^n_p(\delta \vert d) = 0 \, \hbox{ for } p>2. 
\ee 
\end{theorem} 
The proof of this theorem 
is given in \cite{BBH1} and follows from the fact that 
linearized gravity is a linear, irreducible, gauge theory. 
In terms of the characteristic cohomology, this means that all 
conservation laws involving antisymmetric objects of rank $>2$ are 
trivial, $\partial_{\mu_1} S^{\mu_1 \mu_2 \cdots \mu_k} \approx 0 
\Rightarrow S^{\mu_1 \mu_2 \cdots \mu_k} \approx  \partial_{\mu_0} 
R^{\mu_0 \mu_1\cdots \mu_k}$ with $k>2$, $S^{\mu_1 \mu_2 \cdots \mu_k} 
= S^{[\mu_1 \mu_2 \cdots \mu_k]}$, $R^{\mu_0 \mu_1\cdots \mu_k} = 
R^{[\mu_0 \mu_1\cdots \mu_k]}$. [This result holds whether or not  
$S^{\mu_1 \mu_2 \cdots \mu_k}$ carries extra indices.] 
 
In antifield number two, the cohomology is given by the following theorem
(which will be proven below),
\begin{theorem} 
\label{conservation2} 
A complete set of representatives of $H^n_2(\delta \vert d)$  
is given by the antifields $C^{*\m}_a$ conjugate to the 
ghosts, i.e.,  
\be 
\delta a^n_2 + da^{n-1}_1 = 0 \Rightarrow a^n_2 
= \lambda^a_\m C^{*\m}_a dx^0 dx^1 \cdots dx^{n-1} 
+ \delta b^n_3 + d b^{n-1}_2 
\ee 
where the $\lambda^a_\m$ are constant. 
\end{theorem} 
 
In order to interpret this theorem in terms of the characteristic 
cohomology (using Eq.(\ref{redchar0}) and recalling that $n>2$),
 we note that the equations of motion $H^{\m \a}_a = 0$ 
of the linearized theory can be rewritten as 
\be 
H^{\m \a}_a \equiv \partial_\n \Phi^{\m \n \a}_a  
\ee 
with 
\be 
\Phi^{\m \n \a}_a \equiv \partial_\b \Psi^{\m \n \a \b}_a = 
- \Phi^{\n \m \a}_a. 
\ee 
The tensor $\Psi^{\m \n \a \b}_a$ is explicitly given by 
\be 
\Psi^{\m \n \a \b}_a = - \eta^{ \m \a} h^{a \n \b}- 
\eta^{ \n \b} h^{a \m \a}  + \eta^{ \m \b}  
h^{a \n \a} + \eta^{ \n \a} h^{a \m \b} 
+ \eta^{\a \m} \eta^{\b \n} h^a - \eta^{\a\n} \eta^{\b  \m}  h^a 
\ee 
(where $h^a$ is the trace $h^{a \m}_{\; \; \; \m}$) 
and has the symmetries of the Riemann tensor.  The equations of 
motion can thus be viewed as conservation laws involving 
antisymmetric tensors $S^{\m \n}$ 
of rank two, parametrized by further indices 
($\a$ and $a$).  These conservation laws are non-trivial 
because one cannot write $\Phi^{\m \n \a}_a$ as the divergence 
$\partial_\lambda \Theta^{\m \n \lambda \a}_a$ of a tensor 
$\Theta^{\m \n \lambda \a}_a$ that would be completely 
antisymmetric in $\m$, $\n$ and $\lambda$ ($\Psi^{\m \n \a \b}_a$ 
does not have the required symmetries). 
Theorem \ref{conservation2} states that these 
are the only non-trivial conservation laws, i.e., 
\be 
\partial_\nu S^{\mu \nu} \approx  0, \,  
S^{\mu \nu} = - S^{\n \m} 
\Rightarrow  S^{\mu \nu} \approx 
\lambda^a_\a \Phi^{\m \n \a}_a + \partial_\lambda 
U^{\m \n \lambda}, \, U^{\m \n \lambda} = U^{[\m \n \lambda]}. 
\ee   
 
Let us now turn to the proof of Theorem \ref{conservation2}. 
Let $a$ be a solution of the cocycle condition for $H^n_2(\delta |d)$, 
written in dual notations, 
\be 
\delta a + \partial_\m V^\m = 0. 
\label{coca} 
\ee 
Without loss of generality, one can assume that $a$ is linear in 
the undifferentiated antifields, since the derivatives of $C^{* \m}_a$ 
can be removed by integrations by parts (which leaves one in the same 
cohomological class of $H^n_2(\delta |d)$).  Thus, 
\be 
a = f_\m^a C^{* \m}_a + \m 
\label{exprfora} 
\ee 
where $\m$ is quadratic in the antifields $h^{* \m \n}_a$ and their 
derivatives, and where the $f_\m^a$ are functions of $h^a_{\m \n}$ 
and their derivatives.  Because $\delta \m \approx 0$, the 
equation (\ref{coca}) implies the linearized Killing equations for 
$f_\m^a$, 
\be 
\partial_\n f_\m^a + \partial_\m f_\n^a \approx 0. 
\label{linearKilling} 
\ee 
If one differentiates this equation and uses the similar equations obtained 
by appropriate permutations of the spacetime indices, one gets, in 
the standard fashion 
\be 
\partial_\lambda \partial_\n f_\m^a \approx 0. 
\ee 
This implies, using the isomorphism $H^0_0(d |\delta)/R \simeq 
H^n_n(\delta |d)$ and the previous theorem
$H^n_n(\delta |d) = 0$ ($n>2$)
\be 
\partial_\n f_\m^a \approx t_{\m \n}^a 
\label{eqforf} 
\ee 
where the $t_{\m \n}^a$ are constants.  If one splits $t_{\m \n}^a$ 
into  symmetric and antisymmetric parts, $t_{\m \n}^a = s_{\m \n}^a + 
a^a_{\m \n}$, $s_{\m \n}^a = s_{\n \m}^a$, $a_{\m \n}^a = - a_{\n \m}^a$, one 
gets from the linearized Killing equation  
(\ref{linearKilling}) $s_{\m \n}^a \approx 0$ and 
thus $s_{\m \n}^a = 0$ (any constant weakly equal to zero is strongly 
equal to zero).  Let $\bar{f}^a_\m$ be $\bar{f}^a_\m = f_\m^a - a^a_{\m \n} 
x^\n$.  One has from (\ref{eqforf}) 
$\partial_\n \bar{f}^a_\m \approx 0$ and thus, using 
again $H^0_0(d |\delta) \simeq R$, $\bar{f}^a_\m \approx  
\lambda_\m^a$ for some  
constant $\lambda_\m^a$.  This implies $f_\m^a  
\approx \lambda_\m^a + a^a_{\m \n} 
x^\n$: $f_\m^a$ is one-shell equal to a Killing field of 
the flat metric.  If one does not allow for an explicit coordinate 
dependence, as one should in the context of constructing Poincar\'e 
invariant Lagrangians, one has 
$f_\m^a 
\approx \lambda_\m^a$.  
Substituting this expression into (\ref{exprfora}), and 
noting that the term proportional to the equation of motion can be 
absorbed through a redefinition of $\m$, one gets 
\be 
a = \lambda_\m^a  C^{* \m}_a + \m' 
\label{exprforabis} 
\ee 
(up to trivial terms). 
Now, the first term in the right-hand side of (\ref{exprforabis}) is 
a solution of $\delta a + \partial_\m V^\m = 0$ by itself.  This means that 
$\mu'$, which is quadratic in the $h^{* \m \n}_a$ and their derivatives, must 
be also a $\delta$-cocyle modulo $d$.   But it is well known that 
all such cocycles are trivial \cite{BBH1}.  Thus,  
$a$ is given by  
\be 
a = \lambda_\m^a C^{* \m}_a + \hbox{ trivial terms} 
\ee 
as we claimed.  This proves the theorem. 
 
\vspace{.2cm} 
\noindent 
{\bf Comments}  
 
(1) The above theorems provide a complete description of 
$H^n_k(\delta |d)$ for $k>1$. These groups are 
zero ($k>2$) or finite-dimensional ($k=2$).  In 
contrast, the group $H^{n}_1 (\delta |d)$, which is related to 
ordinary conserved currents, is infinite-dimensional since the 
theory is free.  To our knowledge, it has not been completely 
computed.  Fortunately, we shall not need it below. 
 
(2) One can define a generalization of the characteristic cohomology using the 
endomorphism defined in \cite{Dubois-Violette:1999rd}, which 
fulfills $D^3 = 0$ (rather than $d^2 = 0$; 
for more information,  
see \cite{Dubois-Violette:2000ee}).  In the language of 
\cite{Dubois-Violette:1999rd}, the Bianchi identities can be 
written as $D \cdot H = 0$ and follow from the fact that 
$H = D^2 \cdot \Psi$ (just as the Noether identities 
$d M = 0$ for the Maxwell equations $M \approx 0$ follow from 
$M = d ^* F$).  The equations of motion read  
$D^2 \cdot \Psi \approx 0$ and define a non-trivial element 
of a generalized characteristic cohomology involving $D$ rather 
than $d$, since one cannot write $\Psi$ as the $D$ of a local object 
(just as one cannot write $^* F$ as the $d$ of a local object). 
There is thus a close analogy between gravity and the Maxwell 
theory provided one replaces the standard exterior derivative $d$ 
by $D$, and the standard cohomology 
of $d$ by the cohomologies of $D$.   
Note, however, that $\Psi$ is not gauge-invariant, 
while $^* F$ is. 
 
\subsection{Invariant cohomology of $\delta$ modulo $d$} 
We have studied above the cohomology of $\delta$ modulo $d$ in the space 
of arbitary functions of the fields $h^a_{\m \n}$, the antifields, and 
their derivatives.  One can also study $H^n_k(\delta \vert d)$ in the 
space of invariant polynomials in these variables, which involve 
$h^a_{\m \n}$ and its derivatives only through the linearized 
Riemann tensor and its derivatives (as well as 
the antifields and their derivatives). The above theorems remain unchanged 
in this space.  This is a consequence of 
\begin{theorem} 
Let $a$ be an invariant polynomial.  Assume that $a$ is $\delta$ trivial modulo 
$d$ in the space of all (invariant and non-invariant) polynomials, 
$a = \delta b + dc$.  Then, $a$ is $\delta$ trivial modulo 
$d$ in the space of invariant polynomials, i.e., one can assume 
without loss of generality that $b$ and $c$ are invariant polynomials. 
\end{theorem} 
The proof is given in the appendix \ref{A2}.  
 
\section{Construction of the general gauge theory of interacting 
gravitons by means of cohomological techniques} 
\label{hard} 
\setcounter{equation}{0} 
\setcounter{theorem}{0} 
\setcounter{lemma}{0} 

Having reviewed the tools we shall need, we now come to grips with our main
problem: to compute $H^{0,n}(s \vert d)$.
To do this, the main technique is to expand according to the 
antifield number, as in \cite{BBH2}. 
Let $a$ be a solution of  
\be 
sa + db = 0 
\label{S1close} 
\ee 
with ghost number zero. 
One can expand $a$ as 
\be 
a = a_0 + a_1 + \cdots a_k 
\label{expansionofa} 
\end{equation} 
where $a_i$ has antifield number $i$ (and ghost number zero). 
[ Equivalently, $a_i$ has $puregh = i$, i.e. contains $i$'s explicit 
ghost fields $C^a_\a$'s.]
Without loss of generality, one can assume that the expansion 
(\ref{expansionofa}) stops at some finite value of the antifield number. 
This was shown in \cite{BBH2} (section 3), under the 
sole assumption that the first-order deformation of the 
Lagrangian $a_0$ has a finite (but otherwise 
arbitrary) derivative order. 
 
The previous theorems on the characteristic cohomology imply that 
one can remove all components of $a$ with antifield number 
greater than or equal to 3. Indeed, the (invariant) 
characteristic cohomology in degree $k$ measures precisely the obstruction 
for removing from $a$ the term $a_k$ of antifield number $k$ 
(see appendix \ref{A3}).  Since $H^n_k(\delta \vert d)$ 
vanishes for $k \geq 3$ 
by Theorem 4.1, one can 
assume 
\be 
a = a_0 + a_1 + a_2. 
\label{expansionofabis} 
\ee 
Similarly, one can assume (see appendix \ref{A3})
\be 
b = b_0 + b_1. 
\label{expansionofbbis} 
\ee 
 
Inserting the expressions (\ref{expansionofabis}) and  
(\ref{expansionofbbis}) in (\ref{S1close}) we get  
\beq 
\d a_1+\g a_0&=&db_0\label{eq0}\\ 
\d a_2+\g a_1&=&db_1\label{eq1}\\ 
\g a_2&=&0\label{eq2}\,. 
\eeq 
\par 
Recall the meaning of the various terms in $a$ : 
$a_0$ is the deformation of the Lagrangian; 
$a_1$ captures the information about the deformation 
of the gauge transformations; 
while $a_2$ contains the information about the deformation 
of the gauge algebra. We shall first deal with $a_2$, and then 
``descend'' to $a_1$ and $a_0$.
 
\subsection{Determination of $a_2$} 
\label{deterofa2} 
\par 
As we have seen in section \ref{cohoofg}, 
the general solution of (\ref{eq2}) reads, 
modulo trivial terms, 
\be 
a_2 = \sum_J \alpha_J \omega^J 
\ee 
where the $\alpha_J$ are invariant polynomials 
(see (\ref{alphaomega})).  A necessary  
(but not sufficient) condition for  
$a_2$ to be a (non-trivial) solution of (\ref{eq1}), so that $a_1$ exists, 
is that $\alpha_J$ be a (non-trivial) element of 
$H^n_2(\d\vert d)$ (see appendix \ref{A3})
Thus, by
Theorem 4.2, 
the polynomials $\a_J$ must be linear combinations 
of  the 
antifields $C^*_{\a a}$.  
The monomials $\omega^J$ have ghost number two; so they can be 
of only three possible types  
\be 
C^a_{\a}C^b_{\b},~\,\,C^a_{\a}\pa_{[\b}C^b_{\g]},\,\,~ 
\pa_{[\a}C^a_{\b]}\pa_{[\g}C^b_{\d]}. 
\ee 
They should be combined with $C^{*a}_{\a}$ to form $a_2$. 
By Poincar\'e invariance, the only possibility is to take 
$C^a_{\a}\pa_{[\b}C^b_{\g]}$, which  
yields\footnote{Actually, for  
particular values of the dimension $n$, there are 
also solutions of (\ref{eq1}), (\ref{eq2}) built with the $\varepsilon$ 
tensor. If one imposes PT invariance, these possibilities 
are excluded.  Furthermore, they lead to interaction 
terms with three derivatives. 
The corresponding theories will be studied elsewhere \cite{BoulGual}. 
As often in the sequel, we shall 
switch back and forth between a form and its dual without 
changing the notation when no confusion can arise.  So the 
same equation for $a$ is sometimes 
written as $sa + db = 0$ and sometimes written as $sa + \partial_\m 
b^\m = 0$.} 
\be 
a_2=- C^{*\b}_aC^{\a b}\pa_{[\a}C^c_{\b]}a^a_{bc} + \g b_2. 
\label{a2old}
\ee 
Here we have introduced constants $a^a_{bc}$ that parametrize  
the general solution $a_2$  
of equations (\ref{eq1}), (\ref{eq2}). The trivial ``$\g$-exact'' 
additional term in Eq.(\ref{a2old}) will be normalized to a 
convenient value below.
 
The $a^a_{bc}$ can be identified with the 
structure constants of a $N$-dimensional real
algebra ${\cal A}$.  Let $V$ be an ``internal" (real) vector space 
of dimension $N$; we define a product in $V$ through 
\be 
(x \cdot y)^a = a^a_{bc} x^b y^c, \; \; \forall x,y \in V. 
\label{defofproduct} 
\ee 
The vector space $V$ equipped with this product defines the algebra 
${\cal A}$.  At this stage, ${\cal A}$ has no particular further 
structure.  Extra conditions will arise, however, from the demand that 
$a$ (and not just $a_2$) exists and defines a deformation 
that can be continued to all orders.  We shall recover in this manner 
the conditions found in \cite{Wald1}, plus one additional condition 
that will play a crucial role. 
 
It is convenient (to simplify later developments) to fix the $\g$--exact
term in Eq.(\ref{a2old}) to the value
$ b_2 =  \frac{1}{2}C^{*\b}_{a}C^{\a b}h^c_{\a\b}a^a_{bc}$.
Using $ \g h^a_{\a\b} = 2\pa_{(\a}C^a_{\b)}$, we then get,
\beq 
\label{a2} 
a_2=C^{*\b}_{a}C^{\a b}\pa_{\b}C^c_{\a}a^a_{bc}\,. 
\eeq 
In terms of the algebra of the gauge transformations, this term $a_2$ 
implies that the gauge parameter $\zeta^{a \m}$ corresponding to the 
commutator of two gauge 
transformations with parameters 
$\xi^{a \m}$ and $\eta^{a \m}$ is given by 
\be 
\zeta^{a \m} = a^a_{bc} [\xi^b, \eta^c]^\m 
\label{commutatorgaugetransf} 
\ee 
where $[,]$ is the Lie bracket of vector fields. 
It is worth noting that at this stage, we have not used any a priori 
restriction on the number of derivatives (except that it is finite). 
The assumption that the interactions contain at most two derivatives 
will only be needed below.  Thus, the fact that $a$ stops at 
$a_2$, and that $a_2$ is given by (\ref{a2}) 
is quite general. 
 
Differently put: to first-order in the coupling constant, the 
deformation of the algebra of the spin-2 gauge symmetries is 
universal and given by (\ref{a2}).  There is no other possibility. 
In particular, there is no room for deformations of 
the algebra such that the new gauge transformations would 
close only on-shell (terms quadratic in $h^{*}$ are 
absent from (\ref{a2})).  This strengthens the analysis of \cite{Wald1} 
where assumptions on the number of derivatives in the gauge  
transformations were made.
 No such assumption is in fact 
needed.  
 
\subsection{Determination of $a_1$} 
\label{deterofa1} 
\par 
In order to find $a_1$ we have to solve  equation (\ref{eq1}), 
\be 
\d a_2+\g a_1=db_1\,. 
\ee 
We have 
\beq 
\d a_2&=& -2\pa_{\g}h_a^{*\b\g}C^{\a b}\pa_{\b}C^c_{\a}a^a_{bc}= 
- 2\pa_{\g}\ll(h_a^{*\b\g}C^{\a b}\pa_{\b}C^c_{\a}a^a_{bc}\rr)+\nn\\ 
&&2h_a^{*\b\g}\pa_{\g}C^{\a b}\pa_{\b}C^c_{\a}a^a_{bc}+
2h_a^{*\b\g}C^{\a b}\pa_{\b\g}C^c_{\a}a^a_{bc}\,. 
\eeq 
The term with two derivatives of the ghosts is $\gamma$-exact 
(see Eq.(\ref{secondder}), 
thus, for $a_1$ to exist, the term  
$2h_a^{*\b\g}\pa_{\g}C^{\a b}\pa_{\b}C^c_{\a}a^a_{bc}$ 
should be $\gamma$-exact modulo $d$.  But this can happen  
only if is zero. Indeed, we can rewrite it in terms of the generators 
of $H(\gamma)$ by adding a $\gamma$-exact term, as 
\be 
\label{antisymterm} 
2h_a^{*\b\g}\pa_{\g}C^{\a b}\pa_{\b}C^c_{\a}a^a_{bc}= 
2h^{*\b}_{a~~\g }\pa^{[\g}C^{\a] b}\pa_{[\b}C^c_{\a]}a^a_{bc}+ 
\g\ll(\dots\rr). 
\ee 
It is shown in  appendix \ref{mustvanish} that this 
term is trivial only if it vanishes.  Since 
\be 
2h^{*\b}_{~~\g a}\pa^{[\g}C^{\a] b}\pa_{[\b}C^c_{\a]}a^a_{bc} = 
2h^{*\b}_{~~\g a}\pa^{[\g}C^{\a] b}\pa_{[\b}C^c_{\a]}a^a_{[bc]} \, 
\ee 
the vanishing of this term yields 
\be 
a^a_{bc}=a^a_{(bc)}\,, 
\label{commutativity} 
\ee 
namely, the {\it commutativity} of the  algebra ${\cal A}$ defined by the 
$a^a_{bc}$'s.  
This result is not surprising in view of the form of the commutator 
of two gauge transformations since (\ref{commutatorgaugetransf}) 
ought to  be antisymmetric in $\xi^a$ and $\eta^a$.  When (\ref{commutativity}) 
holds, $\d a_2$ becomes 
\be 
\d a_2 =  - 2\pa_{\g}\ll(h_a^{*\b\g}C^{\a b}\pa_{\b}C^c_{\a}a^a_{bc}\rr)+
\gamma\ll(h_a^{*\b\g}C^{\a b}(\pa_{\g}h^c_{\a\b}+\pa_{\b}h^c_{\a\g} 
-\pa_{\a}h^c_{\g\b})a^a_{bc}\rr) 
\ee 
which yields $a_1$ 
\be 
\label{a1} 
a_1= - h_a^{*\b\g}C^{\a b}\ll(\pa_{\g}h^c_{\a\b}+\pa_{\b}h^c_{\a\g} 
-\pa_{\a}h^c_{\g\b}\rr)a^a_{bc}\, 
\ee 
up to a solution of the ``homogenous" equation 
$\gamma a_1 + db_1 = 0$. 
 
As we have seen,  
the solutions of the homogeneous equation do not modify the gauge 
algebra (since they have a vanishing $a_2$), but they do modify 
the gauge transformations.  By a reasoning analogous to the one 
given in the appendix, one can assume $b_1 = 0$ in 
$\gamma a_1 + db_1 = 0$.  Thus, $a_1$ is a $\gamma$-cocycle. 
It must be linear in $h^{* \m \n}_\a$ and in $C^a_\m$ or 
$\partial_{[\m} C^a_{\n]}$.  By Lorentz invariance, it must contain 
at least one linearized curvature since the Lorentz-invariant 
$h^{* \m \n}_\a \partial_{[\m} C^b_{\n]}$ vanishes.  But this 
would lead to an interaction term $a_0$ that would contain at 
least three derivatives and which is thus excluded by our derivative 
assumptions.  Thus, the most general $a_1$ compatible with our 
requirements is given by (\ref{a1}). 
This is the first place where we do need the derivative assumption.  
[We believe that this derivative assumption is in fact not 
needed here in generic spacetime dimensions, if one takes into account 
the other conditions on $a_1$: Poincar\'e invariance, existence of 
$a_0$, etc.  However, we do not have a proof.  More information on this 
in section \ref{derivatives}.]

\subsection{Determination of $a_0$} 
\par 
We now turn to the 
determination of $a_0$, that is, to the  
determination of the deformed Lagrangian at first 
order in $g$. 
The equation for $a_0$ is (\ref{eq0}),  
\be 
\label{eqa0} 
\d a_1+\g a_0=db_0\,. 
\ee 
We have 
\beq 
\d a_1&=&- \frac{\d S_0}{\d h^a_{\a\b}}C^{\g b} 
\ll(\pa_{\a}h^c_{\b\g}+\pa_{\b}h^c_{\a\g}-\pa_{\g}h^c_{\a\b}\rr)a^a_{bc}=\nn\\ 
&&- (\Box h^a_{\a\b}+\pa_{\a\b}h^a-\pa_{\a}\pa^{\r}h^a_{\r\b} 
-\pa_{\b}\pa^{\r}h^a_{\r\a}+\eta_{\a\b}\pa_{\s\r}h^{a \s\r}\nn\\ 
&&-\eta_{\a\b}\Box h^a)C_{\g}^b 
\ll(\pa^{\a}h^{c \b\g}+\pa^{\b}h^{c \a\g}-\pa^{\g}h^{c \a\b}\rr)a_{abc}\,, 
\label{deltaa1} 
\eeq 
where we have defined 
\be 
a_{abc}\equiv k_{ad}a^d_{bc}\,, 
\ee 
where $k_{ab}$ is the quadratic form defined by the free kinetic terms.
Now we prove that (as in Yang-Mills theory) these ``structure constants''
with all indices down, $a_{abc}$, must be {\it fully symmetric},
$a_{abc}=a_{(abc)}$,
for  (\ref{eqa0}) to have a solution. 
\par 
The polynomial $\d a_1$ is trilinear in $\pa_{\a_1\a_2}
h^a_{\a_3 \a_4}$, $\pa_{\a_5} h^b_{\a_6 \a_7}$ and $C^c_{\a_8}$.
There exist twenty-three different ways to contract the Lorentz
indices in the product $\pa_{\a_1\a_2} h^a_{\a_3 \a_4} \,
\, \pa_{\a_5} h^b_{\a_6 \a_7} \, \, C^c_{\a_8}$ to form a
Lorentz scalar.  These are, in full details (and dropping 
the internal indices), 
\beq
\{Q_\Delta\}
&=& \{ \Box h\, \pa_\a h \, C^\a, \Box h \,
\pa^\b h_{\a \b} \, C^\a, \Box h_{\b \g}\,
\pa^\g h^\b_{\; \; \a}\, C^\a, \Box h_{\b \g}\,
\pa_\a h^{\b \g} \, C^\a, \Box h_{\b \g}\,
\pa_\a h^{\b \a} \, C^\g,  \nonumber \\
&& \; \Box h_{\b \g}\, \pa^\b h \, C^\g,
\pa_{\a \b} h^{\a \b} \, \pa_\g h \, C^\g,
\pa_{\a \b} h^{\a \b} \, \pa^\m h_{\m \g}\, C^\g,
\pa_{\a \b} h^{\a \g} \, \pa_\g h^{\b}_{\; \; \m}\,
C^\m, \nonumber \\
&& \;    \pa_{\a \b} h^{\a \g} \, \pa_\m h^{\b}_{\; \; \g} \, 
C^\m, 
 \pa_{\a \b} h^{\a \g} \, \pa_\m h^{\b \m} \,
C_\g, \pa_{\a \b} h^{\a \g} \, \pa^\b h_{\g \m} \, C^\m,
\pa_{\a \b} h^{\a \g} \, \pa^\b h \, C_\g, \nonumber \\
&& \;   
\pa_{\a \b} h^{\a \g} \, \pa^\g h \, C^\b, 
\pa_{\a \b} h^{\a \g} \, \pa^\m h_{\g \m} \, C^\b,
\pa_{\a \b} h \, \pa_\g h^{\a \b}\, C^\g,
\pa_{\a \b} h_{\g \m}\, \pa^\g h^{\a \b}\, C^\m,
 \nonumber \\
&& \;      
\pa_{\a \b} h \, \pa^\b h^\a_{\; \; \g} C^\g,
\pa_{\a \b} h_{\g \m}\, \pa^\b h^{\a \g} C^\m,
\pa_{\a \b} h \, \pa_\g h^{\a \g}\, C^\b, 
\pa_{\a \b} h_{\g \m}\, \pa^\g h^{\a \m} \, C^\b, \nonumber \\
&& \;                                      
\pa_{\a \b} h \, \pa^\a h \, C^\b,
\pa_{\a \b} h_{\g \m}\, \pa^\a h^{\g \m} \, C^\b \}
\eeq
($\Delta = 1, \dots, 23$).  These polynomials are independent:
if $\a^\Delta Q_\Delta = 0$, then $\a^\Delta = 0$; this
can be easily verified.  Consequently,
these polynomials form a basis of the vector space under consideration.
In particular, two polynomials $\a^\Delta Q_\Delta$
and $\b^\Delta Q_\Delta$ are equal if and only if
all their coefficients
are equal, $\a^\Delta = \b^\Delta$.

Let us single out the terms in (\ref{deltaa1}) containing two traces $h^a$; 
there is only one such term, along the first element of the basis, 
\be 
\label{bhcdh} 
- \Box h^a\,\,C^b_{\g}\,\,\pa^{\g}h^c\,a_{abc}\, 
\ee 
By counting derivatives and ghost number, one easily sees that
the solution $a_0$  of  (\ref{eqa0}) must  be a sum of
terms cubic in the fields $h^a_{\a\b}$, with two derivatives
($\gamma$ brings in one derivative). 
The only monomials which give terms with two traces $h^a$ 
by applying the $\g$ operator are 
$h^dh^e\Box h^f$, $h^d\pa^{\m}h^e\pa_{\m}h^f$, 
$\pa^{\m\n}h^d_{\m\n}h^eh^f$, $\pa^{\m}h^d_{\m\n}\pa^{\n}h^eh^f$, 
$h^d_{\m\n}\pa^{\m}h^e\pa^{\n}h^f$ and $h^dh^e_{\m\n}\pa^{\m\n}h^f$. 
Some of these terms are equivalent modulo integrations by parts; only 
three of them are independent, which can be taken to be 
$h^d\pa^{\m}h^e\pa_{\m}h^f$,  
$h^d_{\m\n}\pa^{\m}h^e\pa^{\n}h^f$ and $h^dh^e_{\m\n}\pa^{\m\n}h^f$. 
The piece in $a_0$ that we are considering is then 
\be 
a_0=\dots+h^d\pa^{\m}h^e\pa_{\m}h^f\,b^1_{def}+ 
h^d_{\m\n}\pa^{\m}h^e\pa^{\n}h^f\,b^2_{def}+h^dh^e_{\m\n}\pa^{\m\n}h^f 
\,b^3_{def}\,, 
\ee 
with $b^i_{def}$ being constants with the symmetries  
\be 
b^1_{def}=b^1_{dfe}\,~~b^2_{def}=b^2_{dfe}\,. 
\ee 
Then we apply $\g$ to $a_0$,  
and integrate by parts.  The rationale behind the integrations by 
parts that we perform is to require that the ghosts, which occur 
linearly, should carry no derivatives, as in $\d a_1$. 
Proceeding in this manner and focusing only on 
the terms with two traces $h^a$ in $\g a_0-db_0 = - \delta a_1$, we easily get  
the condition 
\beq 
&&\Box h^aC^b_{\n}\pa^{\n}h^c\ll(4b^1_{cba}-2b^2_{bac}\rr)+ 
C^a_{\m}\pa^{\m\n}h^b\pa_{\n}h^c\ll( 
-4b^1_{abc}+4b^1_{bac}+4b^1_{cab} 
-2b^2_{acb}-2b^3_{cab}\rr)+ \nonumber \\ 
&&h^aC^b_{\n}\Box\pa^{\n}h^c\ll(4b^1_{abc}-2b^3_{abc}\rr)= 
\Box h^aC^b_{\n}\pa^{\n}h^c\,a_{abc}\,. 
\eeq 
{}From this equation, we obtain 
\be 
2b^2_{abc}=4b^1_{cab}-a_{bac}, \; \; 
b^3_{abc}=2b^1_{abc}, \; \; 
-4b^1_{abc}+a_{cab}=0\,. 
\ee 
In particular we find that 
\be 
- a_{abc}=-4b^1_{bca}=-4b^1_{bac}=- a_{cba}\,, 
\ee 
and thus  
\be 
a_{abc} = a_{(abc)} 
\label{allsym} 
\ee   
where we have used the symmetry 
relations of $b^1_{abc}=b^1_{a(bc)}$ and $a_{abc}=a_{a(bc)}$ 
previously derived. 
An algebra which fulfills 
$a_{abc} = a_{cba}$ 
is called Hilbertian, or, in the real case considered here, 
``symmetric". 
 
Now we prove that $a_{abc}=a_{(abc)}$ is a {\it sufficient} condition 
for the (\ref{eqa0}) to have solution. 
This is simply done by explicitly exhibiting a solution. 
Substituting the expression 
\beq 
\label{oura0} 
a_0&=&\ll( 
\frac{1}{4}h^a\pa^{\m}h^b\pa_{\m}h^c 
-\pa_{\m}h^a\pa^{\m}h^{b \a\b}h_{\a\b}^c 
-\frac{1}{4}\pa_{\m}h^{a \a\b}\pa^{\m}h_{\a\b}^bh^c\rr.\nn\\ 
&&\ll.+\pa_{\m}h_{\a\b}^a\pa^{\m}h^{c \b}_{~\g}h^{c \g\a} 
-\pa^{\m\n}h^{a \a\b}h_{\m\n}^bh_{\a\b}^c 
-\frac{1}{2}\pa^{\m}h^{a \a\b}\pa^{\n}h_{\a\b}^bh_{\m\n}^c\rr.\nn\\ 
&&\ll.+\frac{1}{2}\pa^{\m\n}h^ah_{\m\n}^bh^c 
+\frac{1}{2}h^a\pa_{\b}h^{b \b\g}\pa^{\a}h_{\g\a}^c 
-\frac{1}{2}\pa^{\m\n}h^ah_{\; \; \n}^{b \a}h_{\a\m}^c\rr.\nn\\ 
&&\ll.+\pa_{\m}h^ah^{b \m\n}\pa^{\a}h_{\n\a}^c 
-\pa_{\m}h^{a \m\a}\pa_{\n}h^{b\n\b}h_{\a\b}^c 
+h^{a \m\a}h^{b\n\b}\pa_{\m\n}h_{\a\b}^c 
\rr)a_{abc}\nn\\ 
\eeq 
with $a_{abc}=a_{(abc)}$ in the equation (\ref{eqa0}) one finds that it 
is satisfied. The expression (\ref{oura0}) has been derived by considering  
initially the case with one spin two field. In this case, general relativity 
with $g_{\a\b}=\eta_{\a\b}+ g h_{\a\b}$ is a solution 
and the corresponding $a_0$ is the  
term of the Einstein--Hilbert lagrangian cubic in $h_{\a\b}$. 
We verified that this expression satisfies $\d a_1+\g a_0=db_0$, 
and found that the proof remains valid if we take the same expression 
with different fields contracted by a symmetric tensor. 
\par 
We have therefore  proven that a gauge theory of interacting spin two fields, 
with a non trivial gauge algebra, is first-order consistent 
if and only if the algebra ${\cal A}$ defined by $a^a_{bc}$, 
which characterizes $a_2$, 
is commutative and symmetric.

Again, there is some ambiguity in $a_0$ since we can add 
to (\ref{oura0}) any solution of the ``homogeneous" equation 
$\gamma a_0 + db_0 = 0$ without $a_1$.  If one requires that 
$a_0$ has at most two derivatives, there is only one possibility,
namely
\be
- 2 \tilde{\Lambda}_a^{(1)} h^{a \m}_{\; \m}
\ee 
where the $\tilde{\Lambda}_a^{(1)}$'s are constant.  This term
fulfills
\be
\g (\tilde{\Lambda}_a^{(1)} h^{a \m}_{\; \m}) = \partial_\mu (2
\tilde{\Lambda}_a^{(1)} C^{a \mu})
\ee     
and is of course the (linearized) cosmological term.
There is no other non-trivial term.
Indeed, the Euler-Lagrange derivatives $S^{\m \n}
\equiv \delta a_0/ \delta h_{\m \n}$ of
any $a_0$ fulfilling $\gamma a_0 + \partial_\m b_0^\m = 0$ is
an invariant, symmetric tensor fulfilling the contracted Bianchi
identities $\partial_\m S^{\m \n} = 0$ and containing at most
two derivatives.  
Now, the only such tensors 
are $\eta^{\m \n}$ 
and the linearized Einstein tensor.  The first corresponds to 
the cosmological term; the second vanishes on-shell and derives 
from a piece in the Lagrangian that can be absorbed through 
redefinitions of the fields; it is  trivial.

If one does not restrict the derivative order of $a_0$, there 
are further possibilities, e.g., any polynomial in 
the linearized Riemann tensor and its derivatives is a 
solution.  This is the second place where the derivative 
assumption is explicitly used in the analysis.  We shall come 
back to this point in section \ref{derivatives}. 
 
The extra consistency condition  (\ref{allsym}) arises 
because we demand that $a_0$, the first-order 
deformation of the Lagrangian, should exist.  Its 
form explicitly depends on the original Lagrangian 
through the metric $k_{ab}$ defined in 
internal space by the kinetic term. 

The condition (\ref{allsym}) does not appear 
in \cite{Wald1} (although it is mentioned 
in \cite{Anco:1998mf}, but not discussed in the 
context of the free limit).  As we shall see, it is this condition 
that is responsible for the impossibility 
to have consistent cross-couplings between a finite collection of 
(non-ghost) gravitons. 
 
It is interesting to note that a similar phenomenon 
appears in the construction of the Yang-Mills theory from 
a collection of free spin-1 particles.  If one focuses only 
on $a_1$ and $a_2$, one finds that the deformations are characterized 
by a Lie algebra \cite{Wald0}.  But if one requires also that 
$a_0$ exist, the Lie algebra should 
have a further property: it should admit an invariant 
metric, and that metric should be the metric defined by 
the Lagrangian of the free theory (see e.g. \cite{BBH5} 
and references therein).  In the spin-1 case, of course, 
this extra condition does not prevent cross-interactions.

\subsection{The associativity of the algebra from the absence 
of obstructions at second order} 
\par 
The master equation at order two is 
\be 
(W_1,W_1) = - 2 s W_2 
\label{order2} 
\ee 
with 
\be 
W_1 =\int d^nx\,\ll(a_0+a_1+a_2\rr)\,. 
\ee 
One can expand $(W_1,W_1)$ according to the antifield number.  One 
finds 
\be 
(W_1,W_1) = \int d^n x (\a_0 + \a_1 + \a_2) 
\ee 
where the term of antifield number two $\a_2$ comes from 
the antibracket of $\int d^nx \, a_2$ with itself and reads explicitly 
(using (\ref{a2})) 
\be 
\a_2 = -\ll(2C^{*\b}_a\pa_{\b}C^b_{\s}+\pa_{\b}C^{*\b}_aC^b_{\s} 
\rr)C^{\a f}\pa^{\s}C^c_{\a}(a^a_{db}a^d_{fc}). 
\ee  
 
If one also expands $W_2$ according to the antifield number, one gets 
from (\ref{order2}) the following condition on $\a_2$ (it is easy 
to see, by using the arguments given in the appendix, that the expansion 
of $W_2$ can be assumed to stop at antifield number three, 
$W_2 = \int d^n x (c_0 + c_1 + c_2 + c_3)$ and that $c_3$ may be 
assumed to be invariant, $\g c_3 = 0$)  
\be 
\a_2= -2(\g c_2+\d c_3)+db_2\,. 
\ee 
It is impossible to get an expression with three ghosts, one $C^{*\b}_a$
and 
no fields, by applying $\d$ to $c_3$, so we 
can assume without loss of generality that $c_3$ vanishes, 
which implies that  $\a_2$ 
should be $\g$--exact modulo total derivatives. 
\par 
Integrating by parts and adding $\g$--exact terms, one finds 
\be 
\a_2= 
-2 C^{*\b}_a\pa_{[\b}C^b_{\s]}C^f_{\a}\pa^{[\s}C^{\a] c} 
a^a_{d[b}a^d_{f]c} + \hbox{ trivial terms}. 
\ee 
This expression has the standard form (\ref{alphaomega}). It is simple to 
prove,  as in the proof of appendix \ref{mustvanish}, that it  
is not a  mod-$d$ $\g$-coboundary
unless it vanishes.  This 
happens if and only if 
\be 
a^a_{d[b}a^d_{f]c}=0\,, 
\ee 
which is the associative property for the algebra ${\cal A}$ defined by 
the $a^a_{bc}$.  Thus, ${\cal A}$ must be {\it 
commutative, symmetric and associative}. 
 
It is quite important to note that this result holds even if we 
allow more general $a_1$'s or $a_2$'s involving more derivatives, 
since these terms will not contribute to $\a_2$. So, the absence 
of obstructions at order $g^2$ will lead to the same associativity 
condition and the same triviality of the algebra which we 
establish now. 
\par 
\section{Impossibility of cross-interactions} 
\label{obstr} 
\setcounter{equation}{0} 
\setcounter{theorem}{0} 
\setcounter{lemma}{0} 
 
Finite-dimensional real algebras, endowed with a positive-definite 
scalar product, that are commutative, symmetric and associative 
have a trivial structure:  they are the direct sum of one-dimensional 
ideals. 
 
To see this, one proceeds as follows. 
The algebra operation allows us to associate to every element of the algebra 
$u\in\cA$ a linear operator  
\be 
A(u)\,:\,\cA\longrightarrow\cA 
\ee 
defined by 
\be 
A(u)v\equiv u\cdot v\,. 
\ee 
In a basis $(e_1,\dots,e_m)$, one has $v=v^ae_a$ and 
\be 
A(u)^c_{~b}=u^aa^c_{ab}\,. 
\ee 
Because of the associativity property, the operators $A(u)$ provide 
a representation of the algebra 
\be 
A(u)A(v)=A(u\cdot v) 
\ee 
and so, since the algebra is commutative, 
\be 
[A(u),A(v)]=0\,. 
\ee 
 
Now, the free Lagrangian endows the algebra $\cal A$ (viewed as an
$N$-dimensional
vector space) with an Euclidean structure, defined by the scalar product
$(u,v)=k_{ab}u^av^b$. At this point, it is convenient to normalize the 
Euclidean metric $k_{ab}$ in the standard way, $k_{ab}=\d_{ab}$, i.e.
to endow  $\cal A$  with the usual Euclidean scalar product
\be
(u,v)=\d_{ab}u^av^b\,. 
\ee 
The symmetry property  
\be 
a_{abc}=a_{(abc)} 
\ee 
expresses that the operators $A(u)$ are all symmetric  
\be 
(u,A(v)w)=(A(v)u,w)\,, 
\ee 
that is, 
\be 
A(u)=A(u)^T\,. 
\ee 
Then the real, symmetric operators $A(u),~u\in\cA$ are diagonalizable by 
a rotation in $\cal A$, viewed as an
$N$-dimensional Euclidean space .  Since they are commuting,   
they are simultaneously diagonalizable.  In a basis 
$\{e_1,\dots,e_m\}$ in which they are all diagonal, 
one has $A(e_a)e_b =\a(a,b)e_b$ for some numbers $\a(a,b)$ and thus 
\be 
e_a \cdot e_b = A(e_a)e_b=\a(a,b)e_b= e_b \cdot e_a= 
A(e_b) e_a =\a(b,a)e_a\,. 
\ee 
So 
$\a(a,b) = 0$ unless $a = b$.  We set $\a(a,a) \equiv 
\tilde{\kappa}^{(1)}_a$. By using the discrete symmetry 
$ h^a_{\m\n} = -  h^a_{\m\n}$ of the free theory, we can always
enforce that $\tilde{\kappa}^{(1)}_a \geq 0$.

Consequently, the structure constants $a^a_{bc}$ of the algebra 
${\cal A}$ vanish whenever two indices are different.  There is no term in 
$W_1$ coupling the various spin-2 sectors, which are therefore completely 
decoupled.  Only self-interactions are possible.  The first-order 
deformation $W_1$ is in fact the sum of Einstein cubic vertices 
(one for each spin-2 field with $\a(a,a) \not=0$) $+$ (first-order) 
cosmological terms. 
 
Technically, the passage from an arbitrary orthonormal basis in  
internal space 
to the basis where the $A(u)$'s are all diagonal is achieved by 
exponentiating a transformation $\Delta W_1 = (W_1, K_0)$ (see  
(\ref{deltaw1})), where $K_0$ defines an infinitesimal rotation in internal 
space.  It is clear that these rotations leave the free Lagrangian 
invariant ($\Leftrightarrow (W_0,K_0) = 0$).   
So we see that the extra identifications 
of the form $\Delta W_1 = (W_1, K_0)$ have a rather direct and natural 
meaning in the present case. 
 
When none of the $\tilde{\kappa}^{(1)}_a$ vanishes, which is in 
a sense the ``generic case", the basis $\{e_a\}$ is unique. 
The allowed redefinitions $\Delta W= 
(W,K)$ must fulfill 
\be 
(W_0,K_0) = 0, \; \; (W_0,K_1) + (W_1,K_0) = 0 
\ee 
in order to preserve the given $W_0$ and $W_1$. 
The term $(W_1,K_0)$ modify the structure constants $a^a_{bc}$ 
by a rotation and so, cannot be BRST-exact unless it is 
zero.  So, we must have separately $(W_0,K_1) = 0$ and $(W_1,K_0) = 0$. 
Since the basis $\{e_a\}$ in which the $a^a_{bc}$ take 
their canonical form is unique, we infer from $(W_1,K_0) = 0$ 
that $K_0$ is zero.  We can thus conclude that given $W_0$ and $W_1$, 
the redefinition freedom is characterized by a $K = K_0 + g K_1 + 
\cdots$ with $K_0 = 0$ and $(W_0,K_1) = 0$.

\section{Complete Lagrangian} 
\label{complete} 
\setcounter{equation}{0} 
\setcounter{theorem}{0} 
\setcounter{lemma}{0} 
 
With the above information, it is easy to complete the construction of 
the full Lagrangian to all orders in the coupling constant.  This is because 
one knows already one solution , namely the Einstein-Hilbert action.  So, the 
only point that remains to be done 
is to check that there are no others. 
In other words, given $W_0$ and $W_1$, equal to 
the standard Einstein terms, how unique are $W_2$, $W_3$ etc? 
 
One has 
\be 
W = W_0^E + g W_1^E(\tilde{\kappa}^{(1)}_a) 
+ g^2 W_2 + \cdots 
\label{WW} 
\ee 
where we emphasize the dependence of $W_1^E$ on the 
constants $\tilde{\kappa}^{(1)}_a$.
The equation determining $W_2$ is, as we have seen, $sW_2 = -(1/2) 
(W_1^E,W_1^E)$.  A particular solution is the functional 
$W_2^E((\tilde{\kappa}^{(1)}_a)^2)$ corresponding to 
the sum of second-order Einstein deformations, which we know 
exists.  Thus, $W_2 = W_2^E + W'_2$, where $W'_2$ is a solution 
of the homogeneous equation $sW_2' = 0$.  The general 
solution to that equation is $W'_2 =  \tilde{W}_1(b^a_{bc})$, where  
$\tilde{W}_1(b^a_{bc}) \equiv \tilde{W}_1$ has been 
determined in section \ref{hard} and involves at this stage arbitrary 
constants $b^a_{bc}$ fulfilling $b^a_{bc} = b^a_{cb}$ and 
$b_{abc} = b_{(abc)}$. 
 
The equation for $W_3$ is then 
\be 
sW_3 = -(W_2,W_1^E) 
\ee 
i.e., setting $W_3 = W_3^E + W_3'$, where $W_3^E$ is the Einsteinian 
solution of $s W_3^E = -(W_2^E,W_1^E)$, 
\be 
s W'_3 = - (\tilde{W}_1, W^E_1). 
\ee 
Now, $(\tilde{W}_1, W^E_1)$ is $s$-exact if and only if  
the constants $b^a_{bc}$ are subject to 
$a^a_{d[b} b^d_{f]c} +b^a_{d[b} a^d_{f]c} = 0$. 
But this condition expresses that $a^a_{bc} + g b^a_{bc}$ defines 
an associative algebra (to the relevant order). Therefore, 
one can repeat the argument of the previous section: by making an 
order-$g$ rotation of the fields, i.e., by choosing appropriately the term  
$K_1$ in $K$, one can arrange that the only non-vanishing 
components of $b^a_{bc}$ are those with three equal indices, and we set 
$b^a_{aa} = \tilde{\kappa}^{(2)}_a$.  When this is done, we see 
that the term $W_2'$ is equal to $W^E_1(\tilde{\kappa}^{(2)}_a)$ 
and that $W_3'$ is equal to $W_2^E(2 \tilde{\kappa}^{(1)}_a 
\tilde{\kappa}^{(2)}_a)$ 
plus a solution $W''_3$ of the homogeneous equation 
$s W''_3 = 0$.  Continuing in the same way, one easily sees that 
$W''_3 = W^E_1(\tilde{\kappa}^{(3)}_a)$  and the higher 
order terms are determined to follow the same pattern. 
 
Regrouping all the terms in $W$, one finds that $W$ is a sum of 
Einsteinian solutions, one for each massless spin-2 field, with 
coupling constants  
\be 
\kappa_a = g \tilde{\kappa}^{(1)}_a + g^2 \tilde{\kappa}^{(2)}_a + 
g^3 \tilde{\kappa}^{(3)}_a + 
\cdots 
\ee 
For simplicity of notation, we assumed that the cosmological constant
was vanishing at each order.   Had we included it, we would
have found possible cosmological terms for each massless, spin-2 field,
with cosmological constant given by
\be
\Lambda_a = g \tilde{\Lambda}^{(1)}_a + g^2 \tilde{\Lambda}^{(2)}_a +
g^3 \tilde{\Lambda}^{(3)}_a +
\cdots
\ee                 

We can thus conclude that indeed, the most general deformation of the 
action for a collection of free, massless, spin-2 fields is the sum of 
Einstein-Hilbert actions, one for each field, 
\be 
S[g^a_{\mu \nu}] = \sum_a \frac{2}{\kappa_a^2}\int d^nx 
(R^a- 2 \Lambda_a) \sqrt{-g^a}, \; g^a_{\mu \nu} = 
\eta_{\mu \nu} + \kappa^a h^a_{\mu \nu}. 
\label{fullfullHilbert} 
\ee      
as we announced.  There is thus no cross-interaction, to all orders 
in the coupling constants.  
This action is invariant under 
independent diffeomorphisms, 
\be 
\frac{1}{\kappa^a} 
\delta_\epsilon g^a_{\mu \nu} = \epsilon^a_{\m ; \n} +  
\epsilon^a_{\n ; \m} 
\label{fullfull} 
\ee 
and so has manifestly the required number of independent gauge 
symmetries (as many as in the free limit).  
Cosmological terms can arise in the deformation because
they are compatible with the gauge symmetries.
One may view the  
diffeomorphisms (\ref{fullfull})
as algebra-valued diffeomorphisms of a manifold 
of the type considered by Wald \cite{Wald2}, but in the 
present case where the algebra is completely reducible and given by 
the direct sum of one-dimensional ideals, the structure  
of the manifold is rather trivial. 
In the case of a single massless spin-2 field, we recover the 
known results on the uniqueness of the Einstein construction. 
 
If some coupling constants $\kappa^a$ vanish, the corresponding free 
action is undeformed at each order in $g$ and the full 
action coincides, in those sectors, with the free action plus
a possible linearized cosmological term $- 2 \lambda_a h^ {a \m}_{\; \;\;\m}$; 
the gauge symmetry (\ref{fullfull}) reduces of course to the 
original one.  This 
situation is non-generic and unstable under arbitrary deformations. 
By contrast, the Einstein action is stable under arbitrary 
deformations (with at most two derivatives) \cite{BBHgrav}. 

Our no-go theorem generalizes a previous no-go result obtained
by Aragone and Deser, who observed that the coupling of a single massless
spin-2 field  $h_{\mu \nu}$ to an independent
 dynamical metric $g_{\mu \nu}$ was
problematic \cite{ArDe}.  However, our agreement with their
conclusions is interestingly subtle.
The action describing the coupling of a massless spin-2 field
$h_{\mu \nu}$ to a dynamical metric $g_{\mu \nu}$ reads,
to leading order
\begin{equation}
S[g_{\mu \nu},h_{\mu \nu}] = S^E[g_{\mu \nu}] +
S^{PF}[h_{\mu \nu}]_{g} \; ,
\label{begin}
\end{equation} 
where $S^E$ is the Einstein-Hilbert action for $g_{\mu \nu}$
and where $S^{PF}[h]_{g}$ is the Pauli-Fierz action for $h_{\mu \nu}$
in the metric $g_{\mu \nu}$, obtained by replacing ordinary
derivatives by covariant derivatives (minimal coupling).
It is invariant under standard diffeomorphisms,
\begin{equation}
\delta_{\xi}g_{\mu \nu} = {\cal L}_{\xi} g_{\mu \nu} = \xi_{\mu ; \nu}
+ \xi_{\nu ; \mu} \; ; \;
\delta_{\xi}h_{\mu \nu} = {\cal L}_{\xi} h_{\mu \nu} 
= \xi^{\lambda} \partial_{\lambda}h_{\mu \nu} + \partial_{\mu} \xi^{\lambda}
 h_{\lambda \nu} + \partial_{\nu} \xi^{\lambda} h_{\mu \lambda} .
\end{equation}
As it stands, the action (\ref{begin}) fails to be consistent
because it is not invariant under the (covariantized) gauge
symmetries of the massless spin-2 field 
$h_{\mu \nu}$  \cite{ArDe},
\begin{equation}
\delta_{\eta} g_{\mu \nu} = 0, \; \;
\delta_{\eta} h_{\mu \nu} = \eta_{\mu ; \nu} + \eta_{\nu ; \mu}.
\label{backgrtrans}
\end{equation}
Ref.\cite{ArDe} further considered non-minimal couplings of $h_{\mu \nu}$
to $g_{\mu \nu}$, but concluded to the impossibility of curing this
inconsistency. Actually, this problem can be cured:
first (as noted in \cite{ArDe}),
 if one adds to the gauge transformation of the metric
$g_{\mu \nu}$ terms of order $O(h^2)$ (where we
count $\eta^\mu$ as being of order $O(h)$),
one can downplay the original $O(h^2)$-terms in the variation of
the action to the $O(h^4)$ level.  This is because the variation of 
(\ref{begin}) under (\ref{backgrtrans}) is proportional to
the Einstein tensor, i.e., to
the variation of $S^E[g_{\mu \nu}]$ with respect to
$g_{\mu \nu}$ (``first miracle" of \cite{ArDe}).
Second, one can  add higher order
terms (of order $O(h^4)$ and higher)
to the action and modify the
transformation rules for $g_{\mu \nu}$ and $h_{\mu \nu}$
by adding higher-order corrections 
(starting at order $O(h^2)$ and $O(h^3)$, respectively)
in such a way that the complete action is invariant
under the modified gauge symmetries (we differ
here from \cite{ArDe}, who concluded to an obstruction at order
$O(h^4)$).  In fact, the complete consistent action can be easily written
down to all orders.
The result, however, is quite disappointing and reads
\begin{equation}
S[g_{\mu \nu},h_{\mu \nu}] = \frac{1}{2}
\big(S^E[g_{\mu \nu}+h_{\mu \nu}] +
S^E[g_{\mu \nu}-h_{\mu \nu}] \big).
\label{constsum}
\end{equation}        
This action starts like (\ref{begin})
and is manifestly consistent : the full gauge symmetries
are independent diffeomorphisms for $g_{\mu \nu}^{+} \equiv g_{\mu \nu}+h_{\mu \nu}$
and $g_{\mu \nu}^{-} \equiv g_{\mu \nu}-h_{\mu \nu}$ 
and reduce to lowest orders to the above gauge transformations.
However, there evidently exist
variables, namely $g_{\mu \nu}^{+}$ and 
$g_{\mu \nu}^{-}$ in terms of which the cross-interactions
are eliminated and the action reduces to a sum
of standard Einstein actions, in complete agreement with our general
analysis.
 
\section{Infinite-dimensional algebras} 
\label{infinitedim} 
\setcounter{equation}{0} 
\setcounter{theorem}{0} 
\setcounter{lemma}{0} 
 
Our proof above of the absence of cross-interactions between the various massless 
spin-2 fields relied heavily on the fact that all finite-dimensional
associative, commutative and symmetric algebras are trivial.  The proof given 
in Section \ref{obstr}  does not immediately extend to the
 infinite-dimensional case because the operators $A(u)$ may have a continuous
part in their spectrum.  There is, however, a famous isomorphism
theorem in the infinite-dimensional case, due to 
Gelfand and Naimark \cite{GN48}, which
enables one to identify any associative, commutative and symmetric
algebra, endowed with a positive-definite scalar product, to the algebra 
of continuous functions (which vanish at infinity) on some (locally compact)
topological space ${\cal M}$ (constructed as the space of characters of
the algebra). The algebra is realized as the point-wise product of functions,
$ (f.g)(y) \equiv f(y) g(y)$, where $ y \in {\cal M}$, and the scalar
product is the $L^2$ product defined by a measure $ d\mu(y) = \nu(y) dy$
on  ${\cal M}$: $(f,g) = \int \nu(y) dy f(y) g(y)$.
[In full rigour,  the application
of this theorem requires a more precise specification of the 
functional properties of the algebra, but such considerations are
out of place within our purely algebraic approach.]

Once this identification is made, it is easy to see that the consistent
action describing a (possibly uncountably) infinite collection of gravitons
$h_{\mu \nu}(x; y)$ ($y \in {\cal M}$) reads
\begin{equation}
S[g_{\mu \nu}(x; y)] = \int \nu(y) dy \big[ \frac{2}{\kappa(y)^2}\int d^nx
(R(g(x;y)) - 2 \Lambda(y)) \sqrt{-g} \big], 
\label{uncoupled}
\end{equation}
with
\begin{equation}
g_{\mu \nu}(x;y) = \eta_{\mu \nu} + \kappa(y) h_{\mu \nu}(x;y).
\end{equation}
Note that the measure $ d\mu(y) = \nu(y) dy$ can, a priori, contain both
discrete and continuous components, i.e. that the action (\ref{uncoupled}) 
can contain both a series and an integral.
 The curvature $R(g(x,y))$ is computed by treating the $y$'s as  parameters
(one differentiates only with respect to $x$).
Indeed, this action has the correct free field limit and reproduces the
correct cubic vertices $\sim \int d\mu \kappa h \partial h \partial h$.
  It is thus the correct action to all orders
by the formal algebraic extension of the argument used above.  The gauge
symmetries are independent diffeomorphisms in $x$, for each $y$.
The action (\ref{uncoupled}) describes an uncoupled system (one
independent Einstein action at 
each point $y \in {\cal M}$).  Of course, one cannot
find, in general, a countable orthonormal basis 
in which the decoupling is manifest, but (\ref{uncoupled})
shows clearly that the
gravitons live in parallel worlds.  We can conclude that,
even in the infinite-dimensional case, the action can
be rewritten as a sum (integral) of
Einstein actions (provided one can apply the
results of the theorem of Ref.\cite{GN48}). In \cite{Reuter}
a certain Kaluza-Klein model containing an infinite number of
massless gravitons was studied. 
The infinite-dimensional
algebra involved in that model is the algebra
of functions on the round $2$-sphere.  
However, our theorem does not, a priori, apply to this case 
because this model contains an infinite tower of scalar ghosts, whose decoupling
from the spin-two fields has not been established.
 
\section{Coupling to matter} 
\label{mattercoupling} 
\setcounter{equation}{0} 
\setcounter{theorem}{0} 
\setcounter{lemma}{0} 
 
We have shown that a (finite) collection of massless spin-2 fields alone 
cannot have {\it direct} cross-interactions.  One may wonder whether 
the inclusion of matter fields could change this picture: if a given 
matter field was  able to couple to two different gravitons 
simultaneously, we would have, at least, some {\it indirect}
(non local) cross-interactions.  It is of course impossible to 
consider exhaustively all possible types of matter fields. 
We shall consider here only the couplings to a scalar field and 
show that within this framework, cross-interactions remain 
impossible.  Our analysis does not exclude 
possibilities based on a more complicated matter sector, 
but we feel that the simple scalar case is a good illustration of the 
general situation and of the difficulties that should be overcome 
in order to get consistent cross-interactions through matter couplings. 
 
So, we want 
to consistently deform the free theory consisting in $N$ 
copies of linearised gravity plus a scalar field 
\be 
{\cal L}= \sum_a {\cal L}^a_{PF}- \frac{1}{2} \pa ^{\m} \phi \pa _{\m} \phi. 
\ee 
\par 
The BRST differential in the spin-2 sector is unchanged while,
 for the scalar field, it reads
\be 
\g \phi= 0 =  \d \phi, \; \; 
\d \phi^*=\frac{\d S_0}{\d \phi}= 
\Box \phi, \; \; \g \phi^* = 0. 
\ee 
 
Because the matter does not carry a gauge invariance of its own, 
Theorems \ref{vanishing}  
and \ref{conservation2}  on the characteristic cohomology 
remain valid.  This implies that $a_2$ is unchanged and still 
given by  
\be 
a_2= a_2^{\rm old} 
= C^{*\b}_{a} C^{\a b} \pa_{\b} C^c_{\a} a^a_{bc} 
\ee 
even in the presence of the scalar field. 
The scalar field variables can occur only in $a_1$ and $a_0$. 
 
Because $a_2$ is unchanged, $a_1$ will be given by the expression 
found above plus the general solution $\bar{a}_1$ of the  
homogeneous equation $\g \bar{a}_1 + db_1 = 0$, 
\be 
a_1 = a_1^{\rm old}+\bar{a}_1 
\ee 
with  
\be 
a_1^{\rm old} 
=- h^{*\b\g}_a C^{\a b} \left( \pa_{\g}h^c_{\a\b}+\pa_{\b}h^c_{\a\g} 
   -\pa_{\a}h^c_{\g\b}\right)a^a_{bc}.  
\ee 
Without loss of generality, we can assume $\g \bar{a}_1 = 0$ 
(see appendix).  The only possibility compatible with Lorentz-invariance 
and leading to an interaction with no more than two derivatives is 
\be 
\bar{a}_1 = -\phi^* \pa^{\b}\phi C^a_{\b} U^a(\phi) . 
\ee 
Indeed, by integrations by parts, one can assume that no derivative 
of $\phi^*$ occurs, while the term $\pa^{\b}C^a_{\b}$ is $\g$-exact. 
Also, the term $h^{*\a \b}_a C^b_\a \partial_\b V^a_b(\phi)  
\sim - \partial_\b h^{*\a \b}_a C^b_\a V^a_b(\phi) - 
h^{*\a \b}_a \partial_\b C^b_\a V^a_b(\phi)$ is trivial. 
 
Requiring $a_0$ to exist forces the functions $U^a(\phi)$ 
to be constants, so we set $U^a(\phi) = 2 \xi_a$, where 
the $\xi_a$ are constants.  Indeed, in the equation $\d \bar{a}_1 + \g  
\bar{a}_0 + 
\pa_\m k^\m = 0$, one may assume that $\bar{a}_0$ is linear in $h_{\a \b}$ 
since $\bar{a}_1$ is linear in the variables of the gravitational sector 
(ghosts).  One may also assume that $h_{\a \b}$ appears undifferentiated 
since derivatives can be absorbed through integrations by parts.  This yields 
$\bar{a}_0 = h_{\a \b}^a \Psi^{\a \b}_a$ where $\Psi^{\a \b}_a$ involves 
the scalar field and two of its derivatives, $\Psi^{\a \b}_a 
= \pa^\a \phi  \pa^\b \phi P_a (\phi)+ \eta^{\a \b} \pa^\m \phi  
\pa_\m \phi Q_a (\phi) + \pa^{\a \b} \phi R^a (\phi) + \eta^{\a \b} 
\Box \phi S^a (\phi)$, where $P_a (\phi)$, $Q_a (\phi)$, $R^a (\phi)$ 
and $S^a (\phi)$ are some functions of the undifferentiated scalar field. 
Substituting this expression into $\d \bar{a}_1 + \g 
\bar{a}_0 + 
\pa_\m k^\m = 0$ and taking the variational derivative with respect 
to the ghosts gives the desired result $U'^{a} = 0$. 
 
This leads to the following expression for the complete $a_0$, 
\beq 
a_0 &=& a_0^{\rm old} + \bar{a}_0 \\ 
\bar{a}_0 &=& t^{\a\b}h^a_{\a\b}\xi_a 
\eeq 
where $t^{\a\b}$ is the stress-energy tensor of the scalar field 
\be 
t^{\a\b}=\left( \pa^{\a}\phi \pa^{\b}\phi-\frac{1}{2}\eta^{\a\b} 
          \pa^{\m}\phi \pa_{\m}\phi \right). 
\ee                                                          
We thus see that the coupling to the gravitons takes 
the form $t^{\a\b}h_{\a\b}$.  This  
is not an assumption, but follows from the general 
consistency conditions.  Of course, we can also add to 
the deformation of the Lagrangian non-minimal 
terms of the form $V_a(\phi) K^a $,  
which are solutions of the 
``homogeneous equation" $\g a_0 + d b_0 = 0$ without source 
$\d a_1$.  However, such terms vanish on (free) shell and thus 
can be absorbed through field-redefinitions in the adopted 
perturbative scheme.  
 
The previous discussion completely determines the consistent interactions 
to first order.       
In order not to have an obstruction at order 2 in the deformation parameter, 
$(W_1, W_1)$ should be BRST exact.  Now, one has 
\be 
(W_1, W_1) 
  = \int d^n x ((a_1, a_1)+(a_2, a_2)+2(a_0, a_1)+2(a_1, a_2)) 
\ee 
with obvious meaning for the notation $(a_i,a_j)$. 
This should be equal to $-2 s W_2$ and again, without loss 
of generality, we can assume that $W_2$ stops at antifield number 
$2$, $-2 W_2 = \int d^n x(b_0 + b_1 + b_2)$. 
When expanded according to antifield number, 
the condition $(W_1, W_1) = -2 s W_2$ yields (in this precise case)  
\beq 
(a_2, a_2)&=&\g b_2 + d m_2, \\ 
(a_1, a_1) + 2(a_1, a_2) &=& \d b_2 + \g b_1 + d m_1 , \\ 
2 (a_0, a_1) &=& \d  b_1 + \g b_0 + d m_0. 
\eeq 
Taking into account the fact that $a_2 \equiv a_2^{\rm old}$ 
and $a_1^{\rm old}$ fulfill these conditions, one gets the  
following requirement on $\bar{a}_1$ 
\be 
2(\bar{a}_1, a_2)+(\bar{a}_1, \bar{a}_1)= 
\d \tilde{b}_2 + \g \tilde{b}_1 + d \tilde{m}_1, 
\label{aa} 
\ee 
where $\tilde{b}_2$ can be assumed to fulfill 
$\g \tilde{b}_2 = 0$. 
Computing the left-hand side of (\ref{aa}) we get 
\beq 
&&8\pa^{\b}\phi C^a_{\b}\xi_a \pa^{\g}\left( \phi^* C^b_{\b}\xi_b \right) - 
4\phi^*\pa^{\b}\phi C^{\a b}\pa_{\b}C_{\a}^{d}a^c_{bd}\xi_c 
\nonumber \\ 
&&=\pa_{\m}j^{\m}-8\phi^*\pa^{\b}\phi C^b_{\a}\pa^{\a}C^a_{\b}\xi_a\xi_b 
-4\phi^*\pa^{\b}\phi C^{\a b}\pa_{\b}C_{\a}^{d}a^c_{bd}\xi_c. 
\eeq 
Inserting in this expression $\pa_{\b}C_{\a}^{d} = 
\pa_{(\b}C_{\a)}^{d} + \pa_{[\b}C_{\a]}^{d}$, we see that the term 
with symmetrized derivatives is $\g$-exact, while the term 
with antisymmetrized derivatives defines a cocycle of the 
$\g$-cohomology which reads explicitly 
\be 
- 4 \phi^* 2\pa^{[\b}\phi C^{\a]b} \pa_{[\a} C_{\b]}^{b} 
\left( 2\xi_a\xi_b-a^c_{ba}\xi_c \right). 
\ee 
This term is trivial in $H(\g \vert d)$ if and only if its coefficient 
is zero, 
\be 
2\xi_a\xi_b-a^c_{ba}\xi_c = 0 
\ee 
(the term $\d \tilde{b}_2$ contains more derivatives and cannot 
play a  role here). 
In the basis where $a^c_{ba}=0$ if $a\not=b$, one gets $\xi_a\xi_b=0$ when 
$a\neq b$, which means that $\phi$ can couple to only one graviton,
as announced. 
\par

\section{Non positive-definite metric in internal space} 
\label{nondefinitepositive} 
\setcounter{equation}{0} 
\setcounter{theorem}{0} 
\setcounter{lemma}{0} 
 
A crucial assumption in the above derivation of the absence of 
couplings mixing two different massless spin-2 fields was that 
the metric in internal space is positive-definite. This requirement
follows from the basic tenets of (perturbative) field theory, as it is
necessary for the stability of the Minkowski vacuum (absence of 
negative-energy excitations, or of negative-norm states). However, for 
completeness (and for making a link with Ref.\cite{Wald1}), we shall now
formally discuss the case where $\delta_{ab}$ is replaced by
 a non  positive-definite, but still non-degenerate, 
metric $k_{ab}$ in internal space. In this case, the algebra $\cal A$
does not need to be trivial, and
one can construct interacting multi-``graviton'' theories,
 as first shown by Cutler and Wald in the paper 
\cite{Wald1} that initiated our study.  As proven above, these are 
determined by a commutative, associative and symmetric algebra ${\cal A}$
(where ``symmetric'' refers to the condition $a_{abc} = a_{(abc)}$, the 
index $a$ being lowered with a non-positive-definite $k_{ab}$). 
 
As shown in \cite{Wald2}, irreducible, commutative, 
associative algebras can be of either three types: 
\begin{enumerate} 
\item ${\cal A}$ contains no identity element and 
every element of ${\cal A}$ is nilpotent ($v^m = 0$ for 
some $m$). 
\item ${\cal A}$ contains one (and only one) identity element  
$e$ and no element $j$ such that $j^2 = - e$. In that case, 
${\cal A}$ contains a $(N-1)$-dimensional ideal of nilpotent elements 
and one may choose a basis $\{e, v_k \}$ ($k = 1, \cdots, N-1$) 
such that all $v_k$'s are nilpotent. 
\item ${\cal A}$ contains one identity element 
$e$ and an element $j$ such that $j^2 = -e$.  The algebra ${\cal A}$  
is then of even dimension $ N = 2 m$, and there exists a 
$(2(m-1))$-dimensional ideal of nilpotent elements.  One can 
choose a basis $\{e, v_k, j, j \cdot v_k \}$ ($k = 1, \cdots, m-1$) 
such that all $v_k$'s are nilpotent. 
\end{enumerate} 
One can view the third case as a $m$-dimensional complex 
algebra with basis $\{e, v_k \}$. This is  what we shall 
do in the sequel to be able to cover simultaneously both 
cases 2 and 3.  So, when we refer to the dimension,  
it will be understood that this is the complex dimension 
in case 3. 
  
We now show that in cases 2 and 3, the symmetry condition on 
the algebra implies that the 
most nilpotent subspace must be at most one-dimensional. 
This  condition was used in 
\cite{AW} in order to write down Lagrangians. 
 
The most nilpotent subspace of ${\cal A}$ is the subspace of elements 
$x$ that have a vanishing product with everything else, 
except the identity.  More precisely,  
one has  
\be 
e \cdot x = x, \; \; v_k \cdot x = 0. 
\ee 
Let us now compute 
the scalar product $(v_k, x)$.  One has $(v_k, x) = 
(v_k  \cdot e, x) = (A(v_k)e, x)$.  Using the symmetry 
property, this becomes $(A(v_k)e, x) = (e, A(v_k)  x)= 
(e, v_k \cdot x) =  
(e, 0) = 0$.  Thus, one has 
\be 
(v_k, x) = 0, \; \; (e, x)  \not= 0 
\ee 
where the last equality follows from the fact that the 
scalar product defined by $k_{ab}$ must be non-degenerate. 
However, if the most nilpotent subspace has a dimension 
greater than or equal to two, one gets a contradiction 
since if $(e, x_1)= m_1$ and $(e, x_2) = 
m_2$, the non-zero vector $m_2 x_1 - m_1 x_2$ has a vanishing 
scalar product with everything else, implying that $k_{ab}$ 
is degenerate.  QED. 
 
When the most nilpotent subspace is precisely one-dimensional, 
one can write real Lagrangians \cite{Wald1,Anco:1998mf,Wald2,AW}, 
so there exist interacting theories with cross-interactions 
which are consistent from the point of view of gauge 
invariance but which do not have the free field limit 
(\ref{startingpoint}). 
We refer to these works for further 
information.

\section{Analysis without derivative assumptions} 
\label{derivatives} 
\setcounter{equation}{0} 
\setcounter{theorem}{0} 
\setcounter{lemma}{0} 
 
The derivative assumption was used at two places in 
the derivation.  First, in the determination of $a_1$; 
second, in the determination of  
$a_0$.  In both cases, the solution was found to be 
unique only if one restricts the 
number of derivatives. 
 
\subsection{Ambiguities in $a_1$} 
Let us examine first $a_1$.  If one allows more derivatives in $a_1$, 
one can add to $a_1$  terms of the form $\Theta^\a_a C^a_\a 
+ \Theta^{\a \b}_a \pa_{[\a} C^a_{\b]}$ where $\Theta^\a_a$ and 
$\Theta^{\a \b}_a = - \Theta^{\b \a}_a$  
have antifield number one and are annihilated 
by $\g$.   
 
For such additional terms, say $\tilde{a}_1$, to be still compatible 
with the existence of an $a_0$, one must have 
\be 
\d \tilde{a}_1 + \g \tilde{a}_0 + \pa_\m k^\m = 0.  
\label{eqfortab} 
\ee 
One may expand $k^\m$ in derivatives 
of the ghosts as follows, 
\be 
k^\m = t_a^{\m \rho} C^a_\rho + t_a^{\m \rho \sigma} 
\pa_{[\rho} C^a_{\sigma]} + \hbox{ more} 
\ee 
where ``more" contains $\pa_{(\rho} C^a_{\sigma)}$ and higher 
derivatives of $C^a_\rho$.  Using the ambiguity $k^\m \rightarrow 
k^\m + \pa_\n S^{\m \n}$ with $S^{\m \n} = - S^{\n \m}$, one can 
assume $t_a^{\m \rho \sigma} = 0$ (take $S^{\m \n} = \Phi^{\m \n \rho}_a 
C^a_\rho$ and adjust $\Phi^{\m \n \rho}_a = 
- \Phi^{\n \m \rho}_a$ appropriately). 
Substituting this expression for $k^\m$ (with $t_a^{\m \rho \sigma}$ 
equal to zero) in (\ref{eqfortab}) 
yields the following conditions upon equating the coefficients of 
$C^a_\rho$ and $\pa_{[\rho} C^a_{\sigma]}$ (which do not occur in 
$\g \tilde{a}_0$), 
\beq 
\partial_\b t^{\a \b}_a &=& - \d \Theta^\a_a ,\\ 
t^{[\a \b]} &=& - \d \Theta^{\a \b}_a. 
\eeq 
The second of these equations implies that one can get rid of 
$t^{[\a \b]}$ by adding trivial terms.  
 
So we see that the 
interactions defined by the new terms in $a_1$ are determined 
by  symmetric tensors $t^{\a \b}_a$ which are conserved 
modulo the equations of motion ($ \partial_\b t^{\a \b}_a =
 -\d \Theta^\a_a \approx 0$) 
and which are such that $\partial_\b t^{\a \b}_a$ is 
gauge-invariant\footnote{Presumably, this implies that $t^{\a \b}_a$ 
itself can be assumed to be gauge-invariant, so that the corresponding 
interaction is just $h_{\a \b}^a t^{\a \b}_a$.  This interaction 
has the same form as the Einstein self-coupling $h_{\a \b}^a t^{\a \b}_{Ga}$, 
where $t^{\a \b}_{Ga}$ is the energy-momentum tensor of the $a$-th  
massless spin-2 field. 
But neither $t^{\a \b}_{Ga}$ nor $\partial_\b t^{\a \b}_{Ga}$ 
is gauge-invariant.  This is why the Einstein self-coupling   
leads to a non-vanishing $a_2$, i.e., modifies the algebra of the 
gauge transformations.  As we have seen, it is the only coupling 
with this property (up to redefinitions). Note that couplings of 
the form $h_{\a \b}^a t^{\a \b}_a$, with $t^{\a \b}_a$ gauge-invariant 
(if they exist), 
are equivalent to strictly gauge-invariant couplings that do 
not modify the gauge transformations (i.e., are such that $\tilde{a}_1$ can 
be redefined away) if $t^{\a \b}_a = \pa_\m \pa_\n Q^{\a \m \b \n}_a$ 
for some $Q^{\a \m \b \n}_a$ with the symmetries of the Riemann tensor, 
since then $\int h_{\a \b}^a t^{\a \b}_a = \int 
 K^a_{\a \m \b \n} Q^{\a \m \b \n}_a$. 
So we see that the (gauge-invariant) generalized (characteristic) cohomology 
of \cite{Dubois-Violette:1999rd} is also relevant here.}.  
 
Equivalently, in view of Noether's theorem, 
these interactions are determined 
by rigid symmetries with a vector index and an internal 
index, 
\be 
\d_\eta h_{\a \b}^a = \eta_\g^b \Delta^{a \g}_{ b\a \b}([K]) 
\ee 
which commutes with the gauge transformations since the 
coefficients $\Delta^{a \g}_{b \a \b}([K])$ involve the gauge-invariant 
linearized curvatures and their derivatives.  The connection 
between $\Delta^{a \g}_{b \a \b}([K])$ and $\Theta^\g_b$ is simply \cite{BBH1} 
$\Theta^\g_b = h^{*\a \b}_a \Delta^{a \g}_{b\a \b}([K])$.  To 
be compatible with Lorentz invariance, the $\Delta^{a \g}_{ b\a \b}([K])$ 
should transform as indicated by their Lorentz indices.  Furthermore, 
the corresponding Noether charges $t^{\a \b}_b$ should be symmetric in 
$\a$ and $\b$, and two sets of $\Delta^{a \g}_{ b\a \b}$'s that 
differ on-shell by a gauge transformation (with 
gauge parameters involving the curvatures and their 
derivatives) should be identified, since 
they lead to $a_1$'s that differ by trivial terms. 
 
The determination of all the non trivial rigid symmetries with these properties 
(if any) appears to be a rather complicated problem whose resolution 
goes beyond the scope of this paper.  Let us simply point out that 
there exists a similar problem in the case of massless spin-1 fields, 
where these conditions turn out to be so restrictive that they 
admit no non-trivial solution in spacetime dimension $4$ (and presumably 
$>4$ also).  The corresponding problem there is that of determining 
the gauge-invariant conserved currents $j^\m([F])$, which are Lorentz-vectors. 
These lead to interactions of the form $A_\m j^\m$ which do modify 
the gauge transformations but not their algebra ($a_2 = 0$ because $j^\m([F])$ 
is gauge invariant).  Equivalently, one must determine the non-trivial 
rigid symmetries which commute with the gauge transformations and 
the Lorentz transformations.  In 3 spacetime dimensions, there is 
a solution, which yields the Freedman-Townsend vertex (with 
$j_a^\m \sim f_{abc} \epsilon^{\m \a \b} \, ^* \! F^b_\a \, ^* \! F^c_\b$ where 
$\! F^a_\a$ is the 1-form dual to the 2-form $F^a_{\a \b}$) 
\cite{Freed-T,Anco5}. 
In four (and presumably higher) 
spacetime dimensions, there is no solution according 
to a theorem by Torre \cite{Torre}. 
If one believes that the spin-1 case is a good analogy, one would 
expect no non-trivial $\tilde{a}_1$ of the type discussed in this section 
except perhaps in particular spacetime dimensions (furthermore, 
there are further restrictions at order $g^2$ that these $\tilde{a}_1$'s would 
have to satisfy).  If this expectation is correct, the most 
general $a_1$ would be the one given above (subsection 
\ref{deterofa1}), associated 
with the unique $a_2$ determined in subsection \ref{deterofa2}.  
 
\subsection{Ambiguities in $a_0$} 
We now turn to the ambiguity in $a_0$.  Assuming, in view of the 
previous discussion, that $a_1$ is given by (\ref{a1}), we see that 
the most general $a_0$ is given by the particular solution 
(\ref{oura0}) plus the general solution $\tilde{a}_0$ of the equation without 
$a_1$-source 
\be 
\g \tilde{a}_0 + \pa_\m p^\m = 0 \, . 
\label{tildeaO} 
\ee 
The addition of such deformations to the Lagrangian do not 
deform the gauge transformations. 
 
There are two types of solutions to (\ref{tildeaO}): those for which 
$p^\m$ vanishes (or can be made to vanish by redefinitions); and those for 
which the divergence term $\pa_\m p^\m$ is unremovable.   Examples of the 
second type are the cosmological term, the Lagrangian itself and, 
more generally, the leading non-trivial orders of the Lovelock terms  
\cite{Lovelock}.  The first type is given by all strictly gauge-invariant 
expressions, i.e., by the polynomials in the linearized Riemann 
tensors $K^a_{\m \n \a \b}$ and their derivatives 
(without inner contractions since the linearized 
Ricci tensors vanish on-shell and can be eliminated by field redefinitions). 
 
If some of the $a^a_{aa}$'s occuring in $a_2$ vanish, it is clear 
that cross-interactions involving any polynomial in the corresponding 
curvatures  
are consistent to all orders.  If all $a^a_{aa}$'s are not vanishing, 
however, - which is in some sense the ``generic case" -, there 
appear non trivial consistency conditions at order $g^2$.  These conditions 
read 
\be 
(\bar{a}_{0},a_{1})=\g f+dh + \d h. \, . 
\label{equ} 
\ee 
Although we have not investigated in detail this equation for all 
possible $\bar{a}_{0}$'s, we anticipate that it prevents 
cross-terms.  Only terms of the form $\sum_a f_a$ where $f_a$ 
involves only the curvature $K^a_{\m \n \a \b}$ and its derivatives, 
are expected to be allowed.  These lead to consistent interactions 
to all orders, obtained by mere covariantization. 
 
\section{Conclusions} 
\label{conclusions} 
\setcounter{equation}{0} 
\setcounter{theorem}{0} 
\setcounter{lemma}{0} 
 
In this paper, we have established no-go results on cross-interactions 
between a collection of massless spin-2 fields.   
Our method relies on the antifield approach and uses cohomological 
techniques.                                          
 
First, we have shown 
that the only possible deformation of the algebra of the gauge 
symmetries is given by the direct sum of diffeomorphism 
algebras, one in each spin-2 field sector (Eq. (\ref{fullfull}) 
with some $\kappa^a$'s possibly equal to zero).   
This result holds 
independently of any assumption on the number of derivatives 
present in the deformation and  is our main achievement. It goes 
beyond previous studies which restricted the number of derivatives 
in the modified gauge transformations and hence in the modified 
gauge algebra.   
 
Under the assumption that the number of derivatives in the 
interactions  
does not exceed two, we have then derived the most general 
deformation of the Lagrangian, which is a sum of independent 
Einstein-Hilbert actions (with possible cosmological terms 
and again with some $\kappa^a$'s possibly equal to zero), one 
for each spin-2 field (Eq. (\ref{fullfullHilbert})). 
This prevents cross-interactions. Our result formally extends to the
case of an infinite (even uncountable) collection of gravitons.
In that case, the action is an integral over a topological space.
The impossibility to introduce even indirect 
cross-interactions (via the exchange of another sector) remains valid 
if one couples a scalar field (but we have not explored  
all possible matter sectors).  Thus,  
there is only one type of graviton that one can see 
in each ``parallel", non-interacting world. In that 
sense, there is effectively only one massless spin-2 field. The fact that 
the Einstein theory involves only one type of graviton 
is therefore not a choice but a necessity that adds to its 
great theoretical appeal. 
 
We have then discussed how this picture could change if one 
did not restrict the derivative order of the interactions.   
Although the analysis gets then technically 
more involved, we have provided arguments that 
cross-interactions remain impossible (apart from 
the obvious interactions that do not modify the gauge transformations 
and involve polynomials in the linearized curvatures and their  
derivatives).  The only modification appears to be the possible addition 
of higher order curvature terms in each sector.  

Restricted to the case of a single massless spin-2 field,
our study recovers
and somewhat generalizes previous results on the inevitability of
the Einstein vertex and of the diffeomorphism algebra 
by relaxing assumptions usually made on the number of
derivatives in the gauge transformations and on the coupling of matter to
the graviton through the
energy-momentum tensor.
 
The main  virtue of no-go theorems is to put into clear light 
the  assumptions that underlie the negative result under focus.   
In our case, the key assumption is, besides 
locality,
positive-definiteness of the metric in the internal 
space of the gravitons. If this assumption is relaxed, 
cross-interactions become mathematically possible \cite{Wald1}. 
It appears to be difficult, however, to
get a physically meaningful theory because a negative
metric leads to 
negative-energy 
(or negative-norm) states.

\section*{Acknowledgements} 
N.B. and  L.G. are grateful to Glenn Barnich and Xavier Bekaert for  
discussions.  M.H. thanks the Institut des Hautes Etudes 
Scientifiques for its kind hospitality. 
The work of N. B., L.G. and 
M.H. is partially supported by the ``Actions de 
Recherche Concert{\'e}es" of the ``Direction de la Recherche 
Scientifique - Communaut{\'e} Fran{\c c}aise de Belgique", by 
IISN - Belgium (convention 4.4505.86) and by 
Proyectos FONDECYT 1970151 and 7960001 (Chile). 
 
\appendix 
\label{Louies} 
\section{Cohomological results} 
\label{cohomology} 
\setcounter{equation}{0} 
\setcounter{theorem}{0} 
\setcounter{lemma}{0} 
 
The content of this appendix is based on \cite{Louies}. 
 
\subsection{A consequence of Theorem \ref{2.2}} 
\label{A1}
The following useful result follows from Theorem \ref{2.2}. 
If $a$ has strictly positive antifield number, the equation  
\begin{equation} 
\gamma a + d b = 0 
\end{equation} 
is equivalent, up to trivial redefinitions, to 
\begin{equation} 
\gamma a = 0 .
\end{equation} 
That is, one can add $d$-exact terms to $a$, 
$a \rightarrow a' = a  + d v$ such that $\g a'= 0$. 
 
In order to prove this theorem, we consider the 
descent associated with $\gamma a + d b = 0$: from 
this equation, one infers, by using the properties $\gamma^2 = 0$, 
$\gamma d + d \gamma = 0$ and the triviality of the cohomology of $d$, that 
$\gamma b + dc = 0$ for some $c$.  Going on in the same way, we introduce 
a ``descent" $\gamma c + de = 0$, $\gamma e + d f = 0$, etc, 
in which each 
successive equation has one less unit of form-degree. 
The descent ends with last two equations $\gamma m + dn = 0$, 
$\gamma n = 0$ (the last equation is $\gamma n = 0$ either because $n$ 
is a zero-form, or because one stops earlier with a $\gamma$-closed 
term). 
 
Now, because $n$ is $\gamma$-closed, one has, up 
to trivial, irrelevant terms, 
$n = \a_J \omega^J$.  Inserting this into the previous 
equation in the descent yields 
\be 
d (\a_J) \omega^J \pm \a_J d \omega^J + \gamma m = 0 .
\label{keya3} 
\ee 
In order to analyse this equation, we introduce a new differential $D$, 
whose action on $h_{\mu \nu}$, 
$h^{\star}_{\mu \nu}$, $C^{\star}_{\alpha}$ and all their derivatives is the 
same as the action of $d$, but whose action on the ghosts is given by : 
\begin{eqnarray} 
D C_{\mu} &=& {\frac {1}{2}} d x^{\nu} C_{[\mu , \nu ]} 
\nonumber \\ 
D(\partial _{\rho _1 \ldots \rho _s} C_{\mu}) &=& 0 ~ {\rm if}~ s\geq1 .
\end{eqnarray} 
The operator $D$ coincides with $d$ up to $\gamma$-exact terms.  
It follows from the definitions that $D\omega^J = A^J_I \omega^I$ 
for some constant matrix $A^J_I$ that involves $dx^\m$. 
 
One can rewrite (\ref{keya3}) as 
\be 
d (\a_J) \omega^J \pm \a_J D \omega^J + \gamma m' = 0 
\ee 
which implies, 
\be 
d (\a_J) \omega^J \pm \a_J D \omega^J = 0 
\label{keya4} 
\ee 
since a term of the form $\b _J \omega^J $ (with $\b _J$ 
invariant) is $\g$-exact if and only if 
it is zero. 
It is convenient to further split $D$ as the sum of an operator $D_0$ and an  
operator $D_1$. $D_0$ has the same action as $D$ on 
$h_{\mu \nu}$, $h^{\star}_{\mu \nu}$, $C^{\star}_{\alpha}$ and all their 
derivatives, and gives $0$ when acting on the ghosts. $D_1$ gives $0$ when  
acting  on all the variables but the 
ghosts on which it reproduces the action of  
$D$. 
The operator $D_1$ comes with a grading : the number of $C_{[\mu , \nu]}$. 
$D_1$ raises the number of $C_{[\mu , \nu]}$ by one unit, while 
$D_0$ leaves it unchanged.  
We call this grading the $D$-degree.  The $D$-degree is bounded 
because there is a finite number of $C_{[\mu , \nu]}^a$, which are 
anticommuting. 
 
Let us expand (\ref{keya3}) according to the $D$-degree. 
At lowest order, we get 
\be 
d \a_{J_0} = 0 
\ee 
where $J_0$ labels the $\omega^J$ that contain zero derivative of the 
ghosts ($D\omega^J = D_1 \omega^J $ contains at least one derivative). 
This equation implies, according to theorem  \ref{2.2}, that 
$\a_{J_0} = d \b _{J_0}$  where $\b _{J_0}$ is an invariant polynomial. 
Accordingly,  one can write 
\be 
\a_{J_0}\omega^{J_0} = d(\b _{J_0} \omega^{J_0}) \mp 
\b _{J_0} D\omega^{J_0} + \hbox{ $\g$-exact terms}  .
\ee 
The term $\b _{J_0} D\omega^{J_0}$ has $D$-degree equal to $1$. 
Thus, by adding trivial terms to the last term $n$ in the descent, 
we can assume that $n$ contains no term of $D$-degree $0$. 
One can then successively removes the terms of $D$-degree $1$, 
$D$-degree $2$, etc, until one gets $n = 0$.  One then repeats the 
argument for $m$ and the previous terms in the descent until one gets  
$b = 0$, i.e., $\g a = 0$, as requested. 
 
\subsection{Invariant cohomology of $\delta$ modulo $d$.} 
\label{A2}

Throughout this
subsection, there will be no ghost; i.e., the objects that appear
involve only the fields, the antifields and their derivatives.
\begin{theorem}\label{2.6} 
Assume that the invariant polynomial $a_{k}^{p}$ 
($p =$ form-degree, $k =$ antifield number)
is $\delta$-trivial
modulo $d$,
\be
a_{k}^{p} = \delta \mu_{k+1}^{p} + d \mu_{k}^{p-1} ~ ~ (k \geq 2).
\label{2.37}
\ee    
Then, one can always choose $\mu_{k+1}^{p}$ and $\mu_{k}^{p-1}$ to be
invariant.
\end{theorem}

To prove the theorem, we need the
following lemma:
\begin{lemma} \label{l2.1} 
If $a $ is an invariant polynomial that is $\delta$-exact, $a = \d b$,
then, $a $ is $\delta$-exact in the space of invariant polynomials.
That is, one can take $b$ to be also invariant.
\end{lemma} 
{\bf{Demonstration of the lemma}} : 
Any function $f([h], [h^*], [C^*])$ can be viewed as a function
$f(\tilde{h}, [K], [h^*], [C^*])$, where $[K]$ denotes
the linearized curvatures and their derivatives, and where
the $\tilde{h}$ denote a complete set of non-invariant derivatives
of $h^a_{\m \n}$ ($\{ \tilde{h}\} = \{h^a_{\m \n}  ,
\pa_\rho h^a_{\m \n}, \cdots\}$).  (One can put
the $\tilde{h}$ in bijective correspondence with the
ghosts and their derivatives through $\g$.)
The $K$'s are not independent
because of the linearized Bianchi identities, but this does not affect
the argument.
An invariant function is just a function that does not involve
$\tilde{h}$, so one has (if $f$ is invariant), $f = f_{ \vert \tilde{h}=0}$.
Now, the  differential $\d$ commutes with the operation of
setting $\tilde{h}$ to zero.  So, if $a = \d b$ and $a$ is invariant, one
has $ a = a_{ \vert \tilde{h}=0} = (\d b)_{ \vert \tilde{h}=0}
= \d (b_{ \vert \tilde{h}=0})$, which proves the lemma
since $b_{ \vert \tilde{h}=0}$ is invariant.
$\diamond$

{\bf{Demonstration of the theorem}} :
We first derive a chain of equations with the same structure as (\ref{2.37})
\cite{BBH2}.
Acting with $d$ on (\ref{2.37}), we get $d a_{k}^{p} = -
\d d \m^{p}_{k+1}$.  Using the lemma and the fact that
$d a_{k}^{p}$ is invariant, we 
can also write $da_{k}^{p}= -\d a_{k+1}^{p+1}$ with $a_{k+1}^{p+1}$ invariant.
Substituting this in $d a_{k}^{p} = -
\d d \m^{p}_{k+1}$, we get
$\d \left[ a_{k+1}^{p+1}-d \m_{k+1}^{p} \right]=0$. As $H(\d)$ is 
trivial in antifield number $>0$, this yields
\be
a_{k+1}^{p+1}=\d \m^{p+1}_{k+2}+d\m^{p}_{k+1}
\ee
which has the same structure as (\ref{2.37}). We can then repeat
the same
operations, until we reach form-degree $n$,
\be
a^{n}_{k+n-p}=\d \m^{n}_{k+n-p+1}+ d \m^{n-1}_{k+n-p}.
\ee
 
Similarly, one can go down in form-degree.
Acting with $\d$ on  (\ref{2.37}), one gets 
$\d a^{p}_{k}=-d (\d \m^{p-1}_{k})$. If the antifield 
number $k-1$ of $\d a^{p}_{k}$ is greater than or equal
to one (i.e., $k>1$), one can rewrite, thanks to Theorem \ref{2.2},
$\d a^{p}_{k}=-d a^{p-1}_{k-1}$ where $a^{p-1}_{k-1}$ is invariant.
(If $k=1$ we cannot go down and the bottom of the chain is (\ref{2.37})
with $k=1$, namely $a_1^p=\d\m_2^p+d\m_1^{p-1}$.)
Consequently $d \left[ a^{p-1}_{k-1}-\d \m^{p-1}_{k} \right]=0$ and, 
as before, we deduce another equation similar to  (\ref{2.37}) :
\be
a^{p-1}_{k-1}=\d \m^{p-1}_{k}+d\m^{p-1}_{k-1}.
\ee
Applying $\d$ on this equation the descent continues. This descent stops at
form degree zero or antifield number one, whichever is
reached first, i.e.,
\bqn
&{\rm{either}}&~~a^{0}_{k-p}=\d \m^{0}_{k-p+1}
\nonumber \\
&{\rm{or}}&~~a^{p-k+1}_{1}=\d \m^{p-k+1}_{2}+d \m^{p-k}_{1}.
\eqn
Putting all these observations together we can write the entire descent as
\bqn 
a^{n}_{k+n-p}  &=& \d \m^{n}_{k+n-p+1}+d \m^{n-1}_{k+n-p} 
\nonumber \\ 
& \vdots & 
\nonumber \\  
a^{p+1}_{k+1}  &=& \d \m^{p+1}_{k+2}+d \m^{p}_{k+1} 
\nonumber \\
a^{p}_{k}  &=& \d \m^{p}_{k+1}+d \m^{p-1}_{k} 
\nonumber \\
a^{p-1}_{k-1}  &=& \d \m^{p-1}_{k}+d \m^{p-2}_{k-1} 
\nonumber \\
& \vdots & 
\nonumber \\
{\rm{either}}~~a^{0}_{k-p}&=&\d \m^{0}_{k-p+1}
\nonumber \\
{\rm{or}}~~a^{p-k+1}_{1}&=&\d \m^{p-k+1}_{2}+d \m^{p-k}_{1}
\eqn 
where all the $a^{p \pm i}_{k \pm i}$ are invariants.

Now let us show that when one of the $\m$'s in
the chain is invariant,
we can actually choose all the other $\m$'s in such a way that they
share this property. Let us thus assume that $\m^{c-1}_{b}$ is 
invariant. This $\m^{c-1}_{b}$ appears in 
two equations of the descent : 
\bqn
a^{c}_{b} &=& \d \m^{c}_{b+1}+d \m^{c-1}_{b},
\nonumber \\
a^{c-1}_{b-1} &=& \d \m^{c-1}_{b}+ d \m^{c-2}_{b}
\eqn
(if we are at the bottom or at the top, $\m^{c-1}_{b}$ occurs
in only one equation, and one should just proceed from that one).
The first equation tells us that $ \d \m^{c}_{b+1}$ is invariant. Thanks
to Lemma \ref{l2.1} we can choose $\m^{c}_{b+1}$ to be invariant. Looking
at the second equation, we see that $ d \m^{c-2}_{b}$ is invariant
and by virtue of theorem \ref{2.2}, $\m^{c-2}_{b}$ can be chosen
to be invariant since the antifield number $b$ is
positive. These two $\m$'s appear each one in two
different equations of the chain, where we can apply the same reasoning.
The invariance property propagates then to all the
$\m$'s.   
Consequently, it is enough to prove the theorem in form
degree $n$.
   
Now, let us prove the following lemma :
\begin{lemma} \label{l2.2} 
If $a^n_k$ is of antifield number $k>n$, then the ``$\m$"s in (\ref{2.37}) can
be taken to be invariant.
\end{lemma}
{\bf{Demonstration}} : Indeed, if $k>n$, the last
equation of the descent is $a^{0}_{k-n}=\d \m^{0}_{k-n+1}$. We can,
using Lemma \ref{l2.1}, choose $\m^{0}_{k-n+1}$ invariant, and so,
all the $\m$'s can be chosen to have the same
property.$\diamond$  
  
It remains therefore
to demonstrate Theorem \ref{2.6}
in the case where the antifield number
satisfies $k\leq n$. 
Rewriting the top equation (i.e. (\ref{2.37}) with $p=n$)
in dual notation, we have
\be
a_k=\d b_{k+1}+\pa_{\r}j^{\r}_{k},~ (k\geq 2).
\label{2.44}
\ee 
We will work by induction on the antifield number, showing that if 
the property
is true for $k+2$ (with $k>1$), then it is true for $k$. 
As we already know that
it is
true in the case $k>n$, the theorem will be demonstrated.
Let us take the Euler-Lagrange derivatives of (\ref{2.44}).
Since the E.L. derivatives with respect to the 
$C^*_{\a}$ commute with $\d$, we get first :
\be
\frac{\d^Ra_k}{\d C^*_{\a}} =\d Z^{\a}_{k-1}
\label{2.45}
\ee
with 
$Z^{\a}_{k-1}=\frac{\d^Rb_{k+1}}{\d C^*_{\a}}$. For the E.L.
derivatives of $b_{k+1}$ with respect to $h^*_{\m\n}$ we obtain,
after a direct computation, 
\be
\frac{\d^Ra_k}{\d h^*_{\m\n}}=-\d X^{\m\n}_k+2\pa^{(\m}Z^{\n)}_{k-1}.
\label{2.46}
\ee   
where $X^{\m\n}_{k}=\frac{\d b_{k+1}}{\d h^*_{\m\n} }$.
{}Finally,
let us compute the E.L. derivatives of $a_k$ with respect to the 
fields.  We get : 
\be
\frac{\d^Ra_k}{\d h_{\m\n}}=\d Y^{\m\n}_{k+1}-\h^{\m\n}
\pa_{\a\b}X^{\a\b}_{k}-\pa^{\r}\pa_{\r}X^{\m\n}_{k}
+\pa^{\m}\pa_{\r}X^{\r\n}_{k}+\pa^{\n}\pa_{\r}X^{\r\m}_{k}
-\h_{\a\b}\pa^{\m\n}X^{\a\b}_{k}+\h_{\a\b}\h^{\m\n}\pa^{\r}\pa_{\r}
X^{\a\b}_{k}
\label{2.47}
\ee
where $Y^{\m\n}_{k+1}=\frac{\d^Rb_{k+1}}{\d h_{\m\n}}$. 

The E.L. derivatives of an invariant object are invariant. Thus,
$\frac{\d^Ra_k}{\d C^*_{\a}}$ is invariant.  Therefore,
by our lemma \ref{l2.1}
and Eq. (\ref{2.45}), we have also 
\be
\frac{\d^Ra_k}{\d C^*_{\a}} =\d Z'^{\a}_{k-1}
\label{2.45'}
\ee  
for some invariant $Z'^{\a}_{k-1}$.  Similarly, one easily
verifies that 
\be
\frac{\d^Ra_k}{\d h^*_{\m\n}}=-\d X'^{\m\n}_k+2\pa^{(\m}Z'^{\n)}_{k-1}.
\label{2.46'}
\ee   
and
\be
\frac{\d^Ra_k}{\d h_{\m\n}}=\d Y'^{\m\n}_{k+1}-\h^{\m\n}
\pa_{\a\b}X'^{\a\b}_{k}-\pa^{\r}\pa_{\r}X'^{\m\n}_{k}
+\pa^{\m}\pa_{\r}X'^{\r\n}_{k}+\pa^{\n}\pa_{\r}X'^{\r\m}_{k}
-\h_{\a\b}\pa^{\m\n}X'^{\a\b}_{k}+\h_{\a\b}\h^{\m\n}\pa^{\r}\pa_{\r}
X'^{\a\b}_{k}
\label{2.47'}
\ee            
for some invariant $X'^{\m\n}_k$ and $Y'^{\m\n}_{k+1}$.

Now,
since $a_k$ is
invariant, it depends on the fields only
through 
the linearized Riemann tensor and its derivatives.
We can thus write
\be
\frac{\d^Ra_k}{\d h_{\m\n}} = 4\pa_{\a\b}A^{\a\m\b\n}
\label{2.49}
\ee
where 
$A^{\a\m\b\n}$ has the symmetries of
the Riemann tensor.
This implies
\be
\d Y'^{\m\n}_{k+1}=\pa_{\a\b}M^{\a\m\b\n} 
\label{2.50}
\ee
with $M^{\a\m\b\n}$  having the symmetries of the Riemann tensor.
The equation (\ref{2.50}) tells us that the $Y'^{\m\n}_{k+1}$
for given $\n$ are $\delta$-cocycles modulo $d$, in form degree
$n-1$ and antifield number $k+1$. There are thus $\delta$-exact
modulo $d$ ($H^{n-1}_{k+1}(\delta \vert d)
\simeq H^n_{k+2}(\delta \vert d) \simeq 0$),
$Y'^{\m\n}_{k+1}=\d A^{\m\n}_{k+2} +\pa_{\r}T^{\r\m\n}_{k+1}$  where
$T^{\r\m\n}_{k+1}$ is antisymmetric in  $\r$ and $\m$. By our hypothesis of 
induction, $A^{\m\n}_{k+2}$ and  $T^{\r\m\n}_{k+1}$  can be
assumed to be invariant.
Since $Y^{\m\n}_{k+1}$ is symmetric in $\m$ and $\n$,
we have also
$\d A^{[\m\n]}_{k+2} +\pa_{\r}T^{\r[\m\n]}_{k+1}=0$.  The
triviality 
of $H^{n}_{k+2}(d \vert \d)$ implies again
that $A^{[\m\n]}_{k+2}$ and
$T^{\r[\m\n]}_{k+1}$ are trivial, in particular,
$T^{\r[\m\n]}_{k+1}=\d Q^{\r\m\n}_{k+2}+\pa_{\a}S^{\a\r\m\n}_{k+1}$,
where $S^{\a\r\m\n}_{k+1}$ is antisymmetric
in ($ \a, \r$)  and in ($\m,\n$), respectively.
The induction assumption allows us to choose $Q^{\r\m\n}_{k+2}$  and
$S^{\a\r\m\n}_{k+1}$  to be invariant. 
Writing 
$E^{\m\a\n\b}_{k+1}=-[S^{\m\a\n\b}_{k+1}+S^{\n\b\m\a}_{k+1}]$ 
and computing $\pa_{\a\b}E^{\m\a\n\b}_{k+1}$,  we observe finally
that
\be
Y'^{\m\n}_{k+1}=\d F^{\m\n}_{k+2}+\pa_{\a\b}E^{\m\a\n\b}_{k+1}
\label{2.51}
\ee
with $E^{\m\a\n\b}_{k+1} = E^{\n\b\m\a}_{k+1}$,
$E^{\m\a\n\b}_{k+1} = E^{[\m\a]\n\b}_{k+1}$ and
$E^{\m\a\n\b}_{k+1} = E^{\m\a [\n\b]}_{k+1}$.

We can now complete the argument.  Using the homotopy formula
\be
a_k=\int^{1}_{0}dt[\frac{\d^Ra_k}{\d C^*_{\a}}(t)C^*_{\a}+
\frac{\d^Ra_k}{\d h^*_{\m\n}}(t)h^*_{\m\n}+
\frac{\d^Ra_k}{\d h_{\m\n}}(t)h_{\m\n}], 
\label{homotopy}              
\ee 
that enables one to reconstruct $a_k$ from
its E.L. derivatives, as well
as the expressions (\ref{2.45}), (\ref{2.46}), (\ref{2.47}) for
these E.L. derivatives, we get
\be
a_k=\d[\int^{1}_{0}[Z'^{\a}_{k-1}C^*_{\a} 
+X'^{\a\b}_{k}h^*_{\a\b}+Y'^{\m\n}_{k+1}h_{\m\n}]]+\pa_{\r}k^{\r}.
\ee
The first two
terms in the argument of $\d$ are manifestly
invariant.  As to
the third term, we 
use (\ref{2.51}). The $\d$-exact term
disappears ($\d^2=0$) while the second one yields 
$\d [\int^{1}_{0}dt[\pa_{\a\b}E^{\m\a\n\b}_{k+1}h_{\m\n}]]$.
Integrating by part twice gives $E^{\m\a\n\b}_{k+1}$
times the linearized Riemann tensor, which is also invariant.
This proves the theorem.

\subsection{Cohomology of $s$ modulo $d$} 
\label{A3}
We have now developed all the necessary tools for the study of the cohomology of 
$s$ modulo $d$ in form degree $n$. A cocycle of $H(s \vert d)$ 
must obey 
\be 
s a + d b = 0. 
\ee 
Let us expand $a$ and $b$ according to the antifield number : 
\begin{eqnarray} 
a &=& a_0 + a_1 + ...+a_k 
\nonumber \\ 
b &=& b_0 + b_1 + ...+b_l 
\end{eqnarray}   
where, as shown in \cite{BBH2}, the expansion stops at some finite 
antifield number. 
  
Writing $s$ as the sum of $\gamma$ and $\delta$, the equation $s a + d b = 0$ 
is equivalent to the system of equations : 
\begin{eqnarray} 
\delta a_1 + \gamma a_0 + db_0 &=& 0 
\nonumber \\ 
\delta a_2 + \gamma a_1 + db_1 &=& 0 
\nonumber \\ 
& \vdots & 
\nonumber \\  
\delta a_k + \gamma a_{k-1} + db_{k-1} &=& 0 
\nonumber \\ 
 & \vdots & 
\end{eqnarray} 
 
Where the system 
ends depends on $k$ and $l$, but, without loss of 
generality, we can assume that $l = k-1$.  
Indeed, if $l > k-1$ the last equations 
look like $db_i = 0$, (with $i > k$) and imply that (because $b$ is of form  
degree $(n-1)$) $b_i = dc_i$. We can thus absorb these terms in a redefinition of 
$b$. The last equation is then $ \gamma a_k + db_k =0$ which, using  
the consequence of theorem  
\ref{2.2} discussed in appendix
\ref{A1}, can be written $\gamma a_k =0$.  
 
We have then the system of equations (where some $b_i$ could 
be zero): 
  \beq 
  \d a_1+\g a_0 + db_0&=&0 
  \nonumber \\ 
                    &\vdots& 
  \nonumber \\ 
  \d a_k+\g a_{k-1}+db_{k-1}&=&0 
  \nonumber \\ 
               \g a_k &=& 0 .
  \eeq 
The last equation enables us to write $a_k = \alpha_J \omega^J$. 
Acting with 
$\gamma$ on the second to last equation and using $\gamma ^2 =0$~,  
$\gamma a_k = 0$~, we get $d \gamma b_{k-1} = 0$~; and then, thanks to the 
consequence of theorem 
\ref{2.2}, $b_{k-1}$ can also be assumed to 
be invariant, 
$b_{k-1} = \beta _J \omega ^J$. Substituting the invariant forms of 
$a_k$ and $b_{k-1}$ in the second to last equation, we get : 
\be 
 \delta [\alpha_J \omega^J] + D[\beta_J \omega^J] = \gamma (\ldots). 
\label{2.53} 
 \ee 
As above, this 
equation implies 
\be 
\delta [\alpha_J \omega^J] + D[\beta_J \omega^J] = 0. 
\label{2.54} 
\ee 
 
We now expand this equation according to the $D$-degree.  The term 
of degree zero reads 
\be 
[\delta \alpha _{J_0} + D_0 \beta _{J_0}]\omega ^{J_0} =0. 
\ee 
This equation implies that the coefficient of $\omega ^J$ must be zero, and 
as $D_0$ acts on the objects upon which $\beta _J$ depends in the same way as $d$, 
we get : 
\be 
\delta \alpha _{J_0} + d \beta _{J_0} =0. 
\ee 
If the antifield number of $\alpha_{J_0}$ is strictly 
greater than $2$, the solution 
is trivial, thanks to our results on the cohomology 
of $\delta$ modulo $d$: 
\be 
\alpha_{J_0} = \delta \mu_{J_0} + d\nu_{J_0}. 
\ee 
{}Furthermore, theorem \ref{2.6} tells us that  
$\m_{J_0}$ and $\n_{J_0}$ can be chosen 
invariants. We thus get~: 
\beq 
a_{k}^{0}&=&(\d \m_{J_0} +D_0 \n_{J_0})\o^{J_0} 
\nonumber \\ 
&=& s(\m_{J_0}\o^{J_0}) + d (\n_{J_0} \o^{J_0}) + \hbox{ ``more"} 
\eeq 
where ``more" arises from $d \o^{J_0}$ , which can be written as 
$d \o^{J_0} = A^{J_0}_{J_1} \o^{J_1} + su^{J_0}$.  The term  
$\n_{J_0} A^{J_0}_{J_1} \o^{J_1}$ has $D$-degree one, while the 
term $\n_{J_0} s u^{J_0}$ differs from the $s$-exact term 
$s (\pm \n_{J_0}  u^{J_0})$ by  the term $\pm \d(\n_{J_0}) u^{J_0}$, 
which is of lowest antifield number. 
Thus, trivial redefinitions enable one to assume that $a^0_k$ vanishes. 
Once this is done, $\b_{J_0}$ must fulfill $d \b_{J_0} = 0$ and thus 
be $d$-exact in the space of invariant polynomials 
by theorem \ref{2.2} , which enables one to set it to zero 
through appropriate redefinitions. 
 
We can then successively remove the terms of higher $D$-degree 
by a similar procedure, until one has completely redefined away 
$a_k$ and $b_{k-1}$.  One can next repeat the argument for  antifield 
number $k-1$, etc, until one reaches antifield number $2$. 
Consequently, we can indeed assume that the expansion 
of $a$ in Eq. (\ref{S1close}) 
stops at antifield number $2$ and takes the form 
$a = a_0 +a_1 + a_2$ with $b = b_0 + b_1$, as in 
(\ref{expansionofabis}) and (\ref{expansionofbbis}).  Furthermore, the 
last term $a_2$ can be assumed to involve only non-trivial elements 
of the characteristic cohomology $H^n_2(\delta \vert d)$. 
 
\section{Proof of statement made in subsection \ref{deterofa1}} 
\label{mustvanish} 
\setcounter{equation}{0} 
\setcounter{theorem}{0} 
\setcounter{lemma}{0} 
 
We answer in this appendix the question raised in subsection \ref{deterofa1} 
as to whether the term  
$h^{\star \b}_{(a) \g} \pa^{[\g} C^{(b) \a]} \pa_{[\b} C_{\a]}^{(c)} a_{[bc]}^{a}$ 
is $\g$-exact modulo $d$, 
\be 
h^{\star \b}_{(a) \g} \pa^{[\g} C^{(b) \a]} \pa_{[\b} C_{\a]}^{(c)} a_{[bc]}^{a} 
= \g(u) +  \pa_\m k^\m  .
\ee 
Both $u$ and $k^\m$ have antifield number one. 
Without loss of generality, we can assume that $u$ contains 
$h^{\star \b}_{(a) \g}$ undifferentiated, since derivatives can be removed 
through integration by parts. 
As the Euler derivative of a total divergence is zero, we can reformulate the 
question as to whether the following identity holds, 
\be 
 \frac{\d^{L}}{\d  h^{\star \b}_{(a) \g}}  
 ( h^{\star \b}_{(a) \g} \pa^{[\g} C^{(b) \a]} \pa_{[\b} C_{\a]}^{(c)} a_{[bc]}^{a}) 
 = \frac{\d^{L}}{\d  h^{\star \b}_{(a) \g}}(\g u) 
 \ee  
i.e. 
\be 
\pa^{[\g} C^{(b) \a]} \pa_{[\b} C_{\a]}^{(c)} a_{[bc]}^{a} 
= \frac{\d^{L}}{\d  h^{\star \b}_{(a) \g}} \left[ 
  {\rm{~linear ~ combination~of~}} 
 \g          \left\{ \begin{array}{c}   
                  h^{\star} \pa C^{(b)} h^{(c)} 
                \\ h^{\star} C^{(b)} \pa h^{(c)} 
 \end{array} \right\} 
                                           \right]. 
\label{34} 
\ee 
The notations $h^{\star} \pa C^{(b)} h^{(c)}$ and  
$h^{\star} C^{(b)} \pa h^{(c)}$ stand for all terms having these structures. 
 
Now, since $ h^{\star}$ appears undifferentiated in $u$ and hence also in
$\g u$, the Euler-Lagrange derivative with respect to $ h^{\star}$ 
of $\g u$ can be read off straightforwardly and 
is just the coefficient of $ h^{\star}$ in $\g u$, i.e., a 
linear combination of $\g (\pa C^{(b)} h^{(c)})$ and $\g (C^{(b)} \pa h^{(c)})$. 
But none of these terms has the required form to match  
$\pa^{[\g} C^{(b) \a]} \pa_{[\b} C_{\a]}^{(c)} a_{[bc]}^{a}$ since 
$\g (C^{(b)} \pa h^{(c)})$ contains second derivatives of the ghosts 
while $\g (\pa C^{(b)} h^{(c)})$ contains the product of symmetrized 
derivatives with anntisymmetrized derivatives.  This  
establishes the result that  
$h^{\star \b}_{(a) \g} \pa^{[\g} C^{(b) \a]} \pa_{[\b} C_{\a]}^{(c)} a_{[bc]}^{a}$ 
is not $\g$-exact modulo $d$, unless it vanishes.

\end{document}